\documentclass[3p]{elsarticle}

\usepackage{hyperref}

\journal{Advances in Water Resources}

\usepackage{amsfonts}
\usepackage{bm}

\begin{document}

\begin{frontmatter}

\title{Stochastic Longshore Current Dynamics}

\author[mymainaddress]{Juan M. Restrepo\fnref{mycorrespondingauthor} }
\cortext[mycorrespondingauthor]{Corresponding author}
\ead{restrepo@math.oregonstate.edu}
\author[mysecondaryaddress]{Shankar Venkataramani}

\address[mymainaddress]{Department of Mathematics, and College of Earth, Ocean, and Atmospheric Sciences,
Oregon State University,
Corvallis, OR 97331, USA}
\address[mysecondaryaddress]{ Department of Mathematics and  Program in Applied Mathematics,  University of Arizona,
Tucson AZ 85721, USA}

\begin{abstract}

We develop a  stochastic parametrization, based on a `simple' 
 deterministic  model for the dynamics of steady longshore currents, that  produces ensembles  that are statistically consistent  with field observations of these currents.  Unlike deterministic models, stochastic parameterization incorporates randomness and hence can only match the observations in a statistical sense.
 Unlike statistical emulators, in which the model is tuned to the statistical structure of the observation, stochastic parametrization are not directly tuned to match the statistics of the observations. Rather, stochastic parameterization combines 
 deterministic, i.e physics based models with stochastic models for the ``missing physics" to create hybrid models, that are stochastic, but yet can be used for making predictions, especially in the context of data assimilation.

We introduce a novel measure of the utility of stochastic models of complex processes, that we call {\em consistency of sensitivity}. %
A model with poor consistency of sensitivity requires a great deal of tuning of parameters and 
has a very narrow range of realistic parameters leading to outcomes consistent with a reasonable spectrum of physical outcomes. We apply this metric to our stochastic parametrization and show that, the loss of certainty inherent in model due to its stochastic nature  is offset by the model's resulting  consistency of sensitivity. In particular, 
the stochastic model still retains the forward sensitivity of the deterministic model and hence respects important structural/physical constraints, yet has a broader range
of parameters capable of producing outcomes consistent with the field data used in evaluating the model. This leads to an expanded range of model applicability.
We show, in the context of data assimilation, the stochastic parametrization of longshore currents achieves good results in capturing the statistics of observation {\em that were not used} in tuning the model.

\end{abstract}

\begin{keyword}
longshore currents \sep stochastic parametrization \sep parameter sensitivity \sep consistently of model sensitivity \sep data assimilation.
\end{keyword}

\end{frontmatter}


\section{Introduction}

 Longshore (alongshore) currents are ubiquitous oceanic flows in nearshore environments (see \cite{shelfdyns}, for a descriptive review).  
  The two mechanisms responsible for their existence are wave stresses and alongshore sea elevation gradients. 
  Because longshore currents affect nearshore bathymetry and beach morphology, and are responsible for a great deal of nearshore transport, models for these currents  are of great practical utility.

 Presently, deterministic wave-resolving models are used with good results,  capturing some of the complex dynamics of the nearshore, including longshore currents  (see \cite{chen2003}, \cite{choi2015}, \cite{noyes1}, \cite{noyes2}. See also \cite{cienfuegos}). Of note are 
 wave-resolving, depth-integrated models based upon the Boussinesq equations, such as funwaveC (\cite{feddersen14}). These have shown considerable
  forecasting skill (See for example,   \cite{feddersen07}, and \cite{suandafeddersen}, in which field data on eddy variability and dye and drifter dispersion are compared to funwaveC model output). The success of funwaveC   in capturing a wide range of 
 nearshore oceanic phenomena rests upon its generality: it includes higher order dispersion (see \cite{nwogu}), a general bottom drag parametrization (see \cite{feddersen07}),  wave breaking momentum transformation parametrization via Newtonian damping (see \cite{kennedy00}) and  the breaking viscosity
model  \cite{lynett}. 

Non-wave resolving  complex models of the  nearshore ocean environment exist as well. These  also have  compared favorably with certain aspects of longshore current dynamics and observations, such as longshore shear instabilities. They can also capture other nearshore flows, such as rip currents, consistent with observations  ( \cite{jsallen}, \cite{ozkan99}, \cite{UMR} and references contained therein).  
 
None of the models capture longshore current observations in a statistically faithful manner.  By fidelity we mean that the time series generated by the model and the observations are indistinguishable, statistically. The most familiar modeling approach to improving model fidelity is to resolve and include more physics (or better physics). The present state-of-the-art in longshore modeling is the wave-resolving models mentioned above.  An approach that directly focuses on obtaining statistical fidelity is statistical emulation (see  \cite{statem}). 
In this strategy observations 
are used to build phenomenological models. The fundamental modeling strategy consists of proposing  a basic statistical distribution or a regression model. Structure is built into the model by calibrating the model's correlations and other statistical dependences with  data.

The goal of this paper is to propose and demonstrate the use of an alternative modeling approach, called stochastic parametrization. It is an intermediate  between the deterministic and the emulator approaches to modeling physical phenomena. 
This strategy yields a model that has as much structure and physics as possible, leaving as little as possible to chance. A good stochastic parametrization  yields structure in the statistics of its time series by the blend of deterministic and stochastic elements, rather than by tuning the model using data to incorporate the correlations and structure of the observations.

  A common use for stochastic parametrization is to incorporate in a computationally efficient way unresolved dynamics that cannot be ignored even if the  model focused on  large spatio-temporal scale phenomena. 
  A familiar practical example of deterministically parametrizing small scales is via homogeneization (see \cite{homogeneization}): when the small scales offer a certain amount of scale separation and it is statistically homogeneous, the small scale averages that persist appear in the resulting deterministic model
  as complementary or added terms. A more sophisticated approach, in computational fluid mechanics, is  large eddy simulations (LES) of turbulent flows.
  In that case the complementary/added terms themselves have their own dynamics which come from closures of higher moment statistics.
Stochastic parametrization is meant to  increase a model's fidelity, but unlike LES, it will  not do so rationally.  For the longshore current 
model featured in this paper the stochastic parametrization will be used to  capture the small scale variability present in the observations, enhancing this way its fidelity, not its rationality.

In this study we will purposely choose  the simplest possible model for 
longshore current dynamics, a balance model,  as a starting point. Clearly, a model that already has improved physics would be a better choice for the development of an operational stochastic model, but a simple and familiar model makes it plain, to what extent the stochastic parametrization is effective in enhancing the original deterministic  model's fidelity.
 Balance models for longshore currents  capture nothing more than the most basic of physical processes, albeit   under strong assumptions.   Nearly all of the physics in longshore current models are captured by empirical parametrization: the models incorporate parametrized   wave radiation (or the vortex force), wave breaking, turbulence,  stress and drag forces.  
The longshore model we will use is derived from the  vortex force formulation for  the 
evolution of waves and currents in the nearshore (see \cite{MRL04},  and \cite{LRM06}). The vortex force model is a 
 general, non-wave resolving, model.
It was  used in \cite{WULRM} to describe the evolution of rip currents, as well as longshore currents in 
\cite{UMR}.  This    model  (see \ref{model}) will be referred to as the {\it vortex force model}, in order to distinguish it from the simpler and specialized {\it balance model} for longshore current dynamics.

The plan of the rest this paper is as follows:  after describing essential background to the nature of the data, in Section \ref{bdata}, we introduce 
in Section \ref{sec:model} the longshore balance model that will be used as a basis for the stochastic parametrization. 
Section \ref{sec:model}  discusses   the conditions under which the longshore balance model is 
derived  from the non wave-resolving,   wave-current interaction, vortex force model.
 Stochastic parametrization can lead to better   {\it consistency of model sensitivity}, at the expense of increased uncertainty. The topic of consistency of sensitivity will be taken up in Section \ref{sa}. A model that has consistent sensitivity will have a wide range of physically-meaningful parameter combinations with which to capture a  broad spectrum  of  physical outcomes. The balance model will be used to explore and illustrate the consequences of  sensitive consistency.  The stochastic parametrization, inspired by the data and constrained by the physics of longshore currents, is introduced in Section \ref{balance}. Stochastic parametrization of unresolved physics, as evidenced by the data, is used to construct a stochastic balance longshore model. Suggesting a simple model for observational data that is clearly non-Gaussian will lead us to introduce Gaussian mixtures. With this choice of stochastic parametrization the stochastic longshore model  is shown to compare favorably with observations. Notably, the model captures correlations present in the data without having to explicitly put these into the model. Furthermore, 
 the stochastic longshore model will be shown to have  consistent sensitivity.  However, the use of Gaussian mixtures reduces the fidelity of the model. The point of using the mixture model will nevertheless  make the model amenable to simple linear/Gaussian data assimilation methods. Data assimilation is a very useful methodology for combining observations and models. Stochastic models are 
well suited for this application, as will be shown in Section \ref{da}, using the proposed stochastic longshore model and observations in a concrete data assimilation example calculation.

\section{Longshore Current  Observational Data}
\label{bdata}

In the process of constructing a stochastic parametrization, as well as in testing the resulting stochastic model, we will
make use of  field  observations. We will use  the data set collected in
Duck, North Carolina by Herbers, Elgar,  Guza, and Birkemeier, Long and Hathaway  (see \cite{elgar1994}). 
 Henceforth, we shall refer
 to this data as the {\it Duck data}. 
 The Duck data repository  provides data, and in particular,   information on nearshore flow mean velocity. It also has recordings of ocean pressure, temperature, and depth, over the course of several months. It also contains  information on the peak frequency, direction, and sea elevation amplitude
 of incoming waves.  Bathymetry as well as the conditions under which the data was obtained is available as well. The Duck  data as well as ancillary information are  available from:\\
 {\tt frf.usace.army.mil/pub/Experiments/DUCK94/SPUV}.\\

\begin{figure}
\centering
 \includegraphics[scale=0.7]{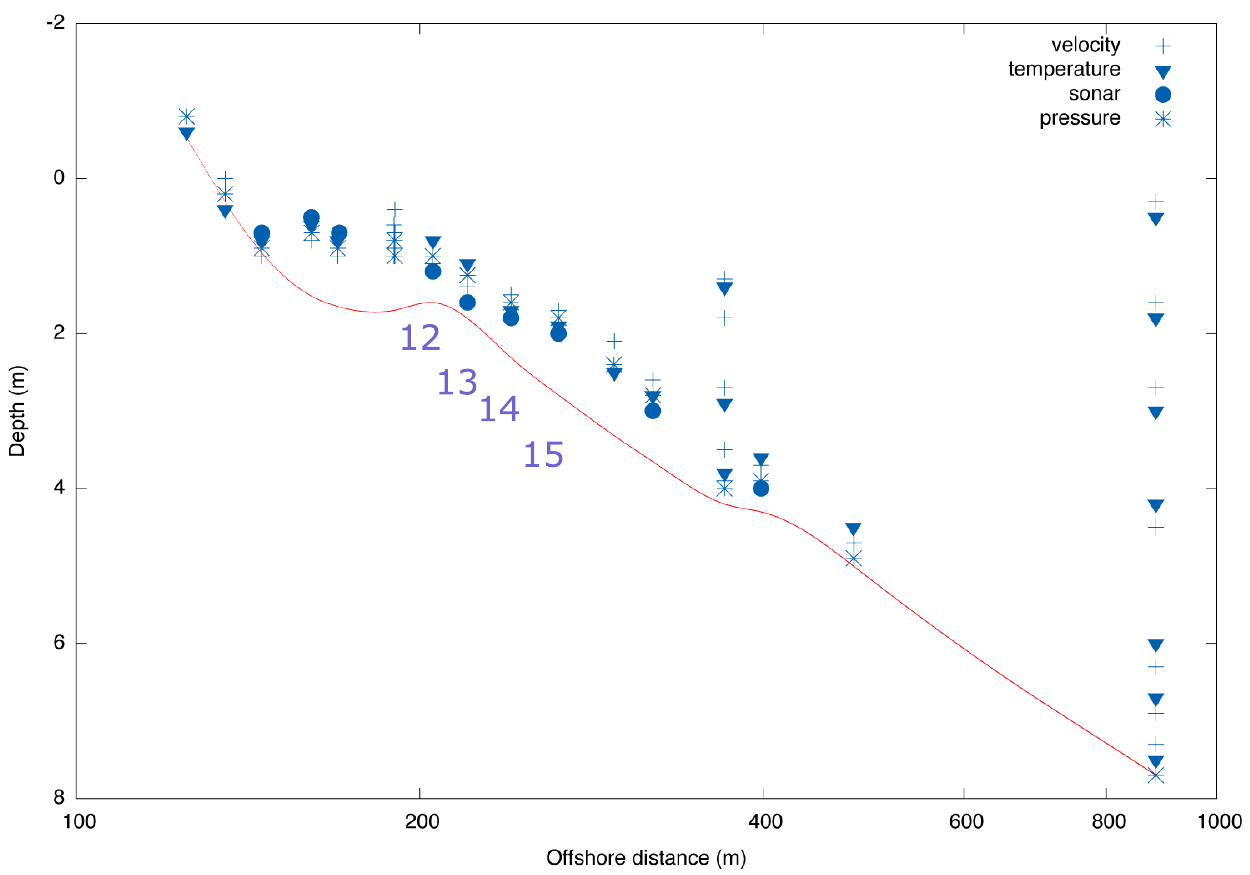}
\caption{Cross section of bathymetry, along the transect $y=929.8$m.  The type of measurements that were collected are indicated.  Stations v12-v15 are indicated as well. }
\label{fig:topography}
\end{figure}

\begin{figure}
\centering
 \includegraphics[scale=0.5]{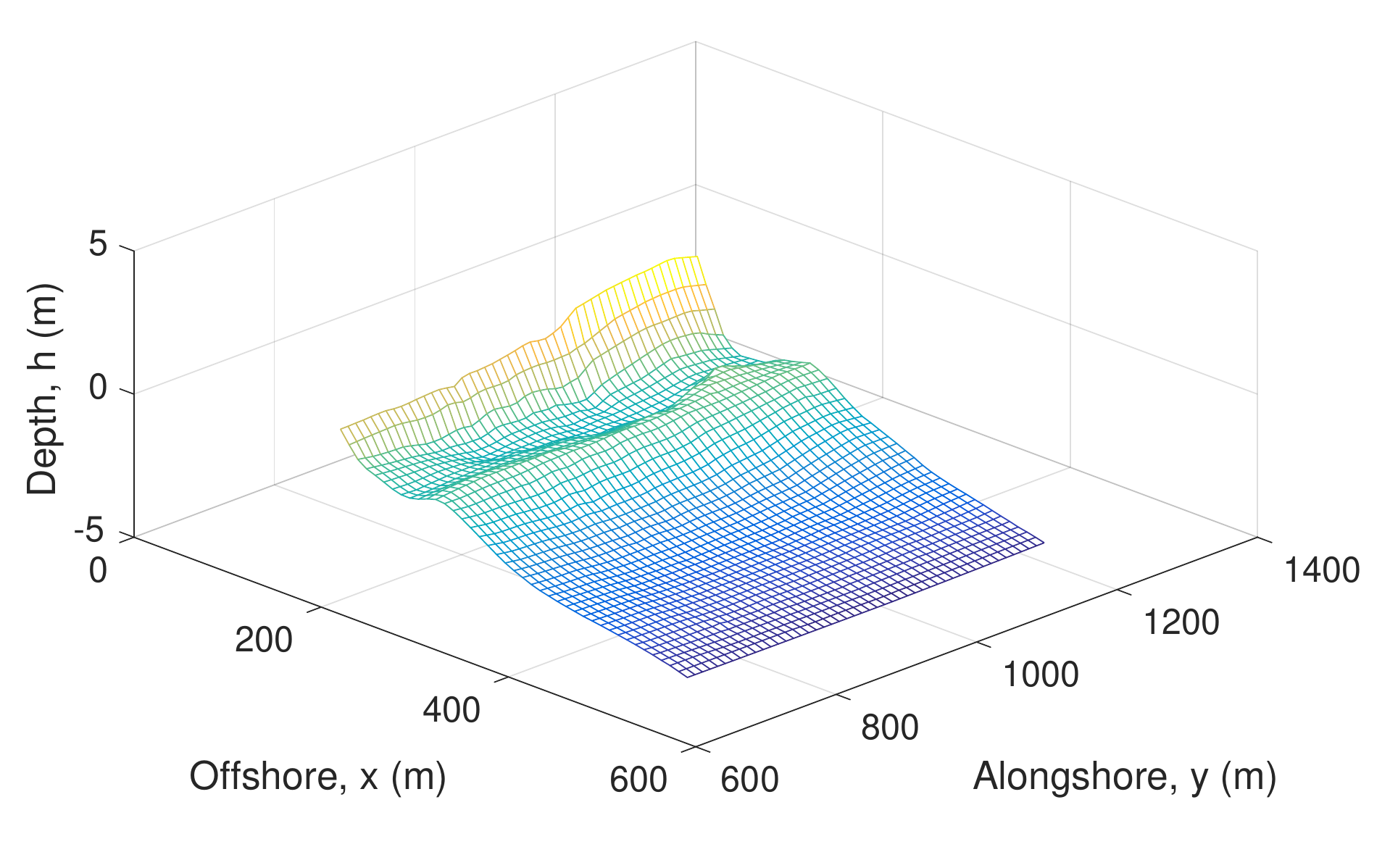}
\caption{The bathymetry, October 1994. The Duck bathymetry is characterized by mild changes in the topography in the 
alongshore direction, compared to changes in the crossshore direction.}
\label{fig:bathy}
\end{figure}

A plot of a  typical  bottom topography cross section appears recreated in Figure \ref{fig:topography}. The plot also shows the data collection devices  and data collection locations. The  cross shore and longshore components of the velocity were collected
at a sampling rate of 2 Hz for 10784 seconds, every 3 hours.  The data collection spanned
 the months of August, September, October, and early
November, 1994. Sporadic instrumentation failure lead to interruptions in data collection. The cross-shore velocity component is zero on average for most of the data gathering campaign, and is ignored in this study.
The specific ''SPUV data'' streams we will make use of are from  
measurement locations v12, v13, v14, v15,
 which were located approximately at the offshore coordinates 205, 220, 240, and 260m, respectively. The transect we will work with is at roughly 930m in the alongshore  ($y$) direction. At these stations the data exhibits good signal to noise characteristics. The locations correspond a location right before 
 the waves shoal and break.
We will also be using 
another set of data, collected during the same time period as the SPUV longshore current data, and it consists of wave elevation, wave period, and wave direction, further out from shore,  roughly 900 m offshore, where the water depth is approximately 8m. This data is
available from 
{\tt frf.usace.army.mil/pub/Experiments/DUCK94/FRF}. 
Figure  \ref{fig:bathy} shows a more general view of the bathymetry, reconstructed from October 1994 data. As can be seen, 
the alongshore variability in the bathymetry is  mild and is used to justify picking a single $y$-location for the data. 
In any event, our focus is on the methodology of stochastic parametrization, rather than on observations or the validity of the 
specific model chosen.

 \section{The Longshore Current Balance Model}
 \label{sec:model}
 
One of the earliest balance models for longshore currents is due to Longuet-Higgins  \cite{LH70, LH70b}. In his model longshore currents are 
driven by the wave radiation stress  (see \cite{LHS60,LHS61}). Several several simplifying assumptions were made in the formulation of the model: the current was assumed steady,   the bottom $h(x)$ was assumed to be gently sloping,  {\it i.e.}, with a slope $\beta := dh/dx$, where $x$ is the (offshore) distance from the shore, in the range  $0 < \beta  \ll 1$. The bottom was also assumed  featureless, and void of  alongshore $y$ dependence. Nonlinear effects were ignored, wave refraction was ignored and the angle of incidence of the waves was assumed to not differ considerably from the normal to the beach. Finally,  the water column depth $H$ was approximated by the depth $h$, which is  the distance from the bottom to the quiescent sea level.

 In its classic form the model suggests that longshore currents result  from  a balance of the drag force ${\bf D}$,  the gradient of the (transverse) radiation stress tensor $S$ due to waves, and dissipation ${\bf N}$.  With the depth-averaged and time-averaged current denoted by ${\bf u}=(u,v)$,  the  current $v$ is given by the $y$-component (alongshore) of the momentum balance. The balance reads
\begin{equation}
0 = -c_D \, v  - \frac{\partial S_{xy}}{\partial x} + N  \frac{\partial}{\partial x} \left( \sqrt{g h} h \frac{\partial v}{\partial x} \right).
\label{lh}
\end{equation}
The first term on the right hand side on (\ref{lh}) is the  bottom drag, with a particularly simple parametrization: the drag force is proportional to the current, $c_D \ge0$ is the proportionality constant. The second term is the  gradient of the net stress
per unit area  due to waves, where $S_{xy} \approx  \sigma W   \cos \theta \sin \theta$. $W$ is the wave action, $\sigma$ is the wave frequency. The angle $\theta$ is with respect to the coordinate $x$, normal to the shore. The last term is the $y$-component of the  lateral dissipation, $N$ is a dimensionless tunable parameter.  This particular dissipation model  was suggested by  Longuet-Higgins, \cite{LH70, LH70b}. It is  based upon dimensional arguments and makes many approximations including the assumption that the   longshore currents occur over  mildly sloped
bathymetry. (See \cite{ruessink}, for an analysis and  discussion of this dissipation model and some of its alternatives).

In what follows we will use a similar and equally simple balance model. We will denote it the {\it Longshore Balance Model}, hereon.
The model derives from the {\it Vortex Force} model for wave current interactions, which is described in 
 \ref{model}.  Similar assumptions as those made by Longuet-Higgins take us from the vortex force model to the longshore balance model  (see discussion leading to (\ref{longshore}). The longshore model is
\begin{equation}
0=-c_D \, v +\alpha  \frac{\beta_B }{h^{5}} +N   \frac{\partial}{\partial x}\left(\sqrt{g h} h  \frac{\partial v}{\partial x} \right).
 \label{eq:LongHig2}
\end{equation}
In (\ref{eq:LongHig2}) 
\[
\alpha := 12/\sqrt{\pi} g B_r^3/\gamma^4 \ge 0, \quad \quad
\beta_B:= A^7 k \sin(\theta).
\]
  The wavenumber magnitude is $k$.  
the sea elevation is $A$, and $g$ is gravity.  
 $B_r$ and $\gamma$ are parameters associated with wave breaking and sea elevation, respectively.
 
The longshore balance model  is equivalent to a balance model proposed by Thornton and Guza \cite{TG83}.  When they compared their model to
field data, obtained  off the gently sloping coast of  their test site in California (USA), they found good agreement. This agreement seemed not to depend in any strong way on having dissipation. However,  when they  compared their model to data obtained over a barred beach, such as the Duck site data, 
their model was inadequate. Further analyses of  comparisons of models and the Duck data associated with the   DELILAH  field campaign \cite{chth93} 
brought into question  the linear model for the bottom drag force, and further, the need for  a spatially varying bottom friction in order to get better agreement between model and data.  Over the years several improved models
for the bottom drag have been proposed. See, for example,  \cite{feddersen2000}, for a bottom drag model that is directly inspired by the Duck and SuperDuck data sets.
Nevertheless, we will keep the linear bottom drag model in the longshore balance model, since it leads to special modeling challenges and the goal  of this work 
is to demonstrate  stochastic parametrization as a tool with which to improve model fidelity.

 \section{Consistency of Sensitivity Analysis}
\label{sa}

There are several practical reasons for an analysis of the model's sensitivity, to either forcing/boundary conditions, initial conditions, or parameters. One reason is to identify which of these has the most impact
(usually in the linearized sense) on the model outcomes (see \cite{ser}). 
Sensitivity and uncertainty are two different things, but they are sometimes intertwined. Another reason is to 
diagnose sensitive dependence on initial conditions  in evolution equations. Sensitive dependence is a hallmark of chaotic systems. Yet another reason
is to evaluate the robustness of  numerical approximations to evolution equations. 

The most common way to assess these sensitivities is by a forward linearized approach. 
In  forward sensitivity analysis we want to determine how relative perturbations of the outputs depend on relative perturbations of the inputs (in backward sensitivity analysis we instead ask what relative inputs are required to produce a certain relative perturbation of the output). 

In similar fashion, one could also determine explicitly or implicitly the functional relationship between model outcomes and perturbations of 
parameters.  One outcome is to determine whether a broad spectrum of physical outcomes are reached by reasonable and sensible range of 
parameters. How large should one determine the ranges of outputs and input parameters? One reasonable approach is to seek similar ranges in 
the physics of its measurements and the physical variables that inform the parameters in the model. Consistency of sensitivity (to parameters) is similarity in the structure and the relative magnitudes of variations in the model outcomes to variations in the parameters, and their measured values. Clearly, the concept is qualitative and it is not universally applicable: some model parameters have no physical counterpart.

Forward (linear) sensitivity analysis is used to determine the explicit dependency of longshore velocity fluctuations on the wave forcing. 
These fluctuations are with respect to the ensemble mean, denoted by  $\langle \cdot  \rangle$. In what follows we will ignore the lateral dissipation, as it is thought to be less critical  to longshore dynamics than the other forces in the longshore balance model (see \cite{chth93}).
Ignoring dissipation in (\ref{eq:LongHig2}) we get the balance equation
\begin{equation}
\frac{c_D }{h} \langle v\rangle = \alpha \frac{\langle k \sin \theta A^7 \rangle}{h^6},
\label{eq:aim2}
\end{equation}
The sensitivity of the longshore current  is obtained by taking the first variation to obtain
\begin{equation}
 \frac{c_D}{h} \langle \Delta v \rangle +  \frac{\Delta c_D}{h} \langle v \rangle  \approx  \left (\frac{\Delta k}{k} + \frac{7 \Delta A}{A^7} + \frac{\cos \theta \Delta \theta}{\sin \theta} 
\right) \langle v \rangle.
\label{eq:aim1}
\end{equation}
The sensitivity in the angle,
even when small, can be physically dramatic when the average angle of incidence is nearly zero since $\sin(\theta)$ in the denominator is small.
Nevertheless, we will focus in this study on the variability of velocity fluctuations to changes in $A$ and $c_D$.
Concerning these,
\[
\langle\Delta v\rangle\propto  \frac{k \sin \theta A^7}{h^5 c_D}\left(\frac{7 \Delta A}{A} - \frac{\Delta c_D}{c_D}\right).
\]
We denote  the wave amplitude supplied by wave forcing from the deep waters as $A_\infty$. 
Since $A=A(A_\infty)$ (see ahead, in  (\ref{aeq}), for the specifics), we approximate $\Delta A$ by $\Delta A_\infty$, and assume that the relative variability in the amplitude $\Delta A/A \gg \Delta c_D/C_D$, the relative variability in the drage coefficient. With these approximations,
\begin{equation}
\langle\Delta v\rangle\approx\mathrm{const}\frac{k \sin \theta}{h^5}\frac{\Delta A_\infty}{c_D}.
\label{dv}
\end{equation} 
We refer to  (\ref{dv}) as  the drag-wave forcing  {\it sensitivity analysis estimate}, and we will 
build a stochastic parametrization that respects, to a certain extent, this analysis.

\subsection{Consistency of sensitivity analysis for the Deterministic  Model}
\label{testing}

There are two challenges in using data to  tune model parameters. Posed as questions,  what sort of statistics do we apply on the data to affect the comparison to the model? And the more challenging question, for a given set of parameters, do we 
retain consistency  with the inherent structure of the model, for a reasonable range of model inputs?

The Duck data will be used to tune the model represented by (\ref{eq:LongHig2}). (The dissipation will be ignored).  The entire 8m depth offshore   time series data set 
for  $k$, $\theta$, and $A_\infty$ is used to get estimates of their mean values. (The dispersion relationship (\ref{dispersion}) and wave number conservation  (\ref{eq:wavenumbers}) equations are required to relate the observed wave period to $k$). The wave action equation (\ref{eq:waveaction}) is needed to relate $A$ at the measuring stations  to the observational data, $A_\infty$, obtained in the deeper waters. We then proceed by finding $c_D$ so that the mean data for $v$ at station v13  coincides with the predicted value of $v$, via (\ref{eq:LongHig2}), at that location.

\begin{figure}
\centering
\includegraphics[height=2in,width=3in]{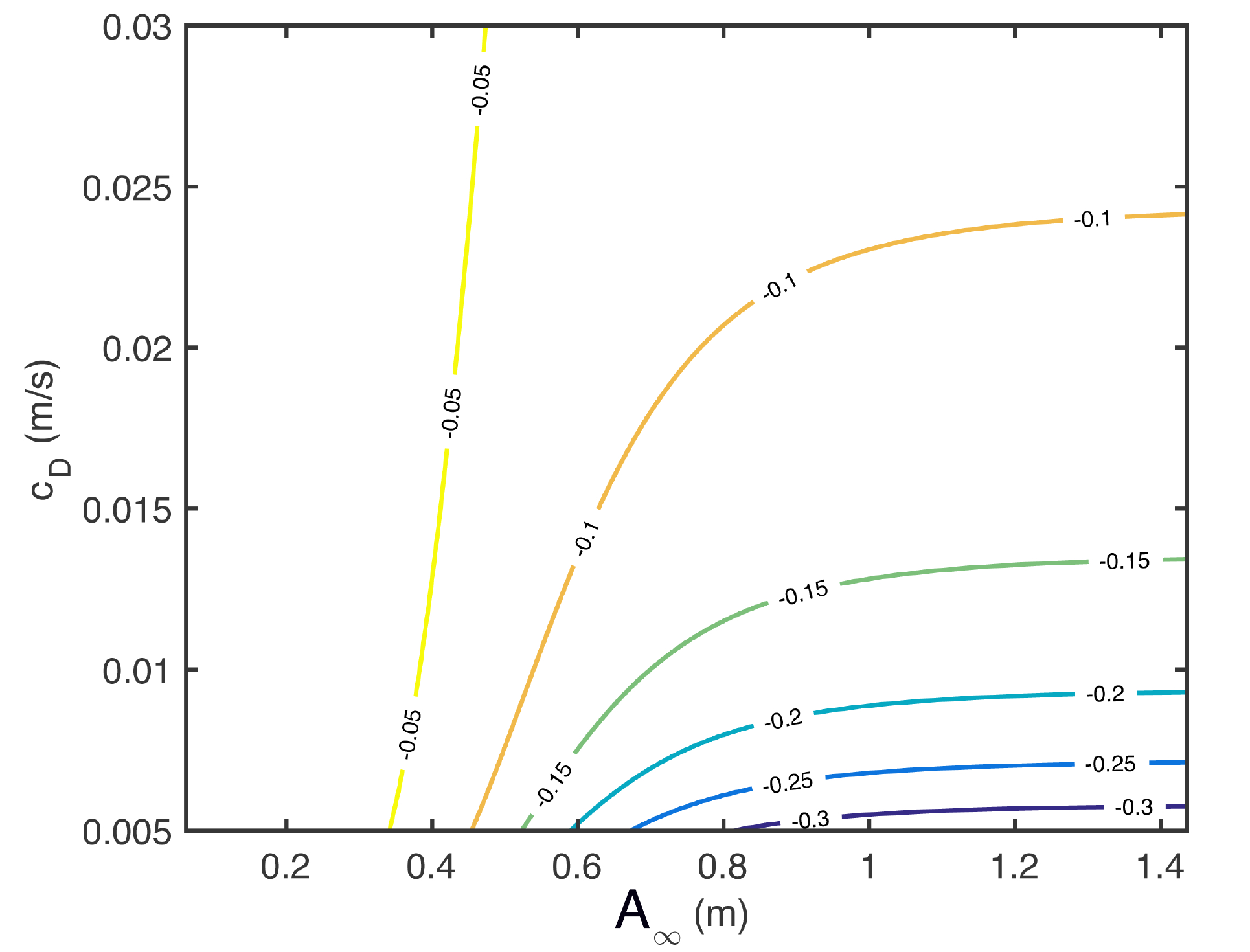}
\caption{Sensitivity curves for  the  balance model,  (\ref{eq:LongHig2}), in $(A_\infty,c_D)$ parameter space.  The model was forced using observational data captured in  8m depth offshore waters. For the wave forcing $a$ a time average value is used. Time average values were also used for 
$\theta$ and for $k$.  The contours represent the mean of the difference between the estimated velocity and the measured velocity at station v13.}
\label{fig:estd}
\end{figure}

 Figure \ref{fig:estd} shows contours of discrepancy between the value of the calculated and observed longshore velocity at station v13.  
We note that for small off shore wave amplitudes $A_\infty$, the longshore velocity is relatively insensitive to the drag parameter, while for large $A_\infty$, the longshore velocity is indeed sensitively dependent on $c_D$.
The conclusion from this exercise is that for certain data one is forced to choose unphysical  combinations of $(A_\infty,c_D$)  to obtain agreement between model outcomes and field data, or parameter choices that make the model unacceptably sensitive. Both of these issues are symptomatic of inconsistency of the sensitivity:   even if one obtains reasonable agreement between the model and the data some of the combinations that lead to agreement between model and data may not respect the sensitivity 
estimate,  (\ref{dv}). (Fortunately the situation is less dramatic when  the (full) complex model is used instead and is tuned to obtain reasonable agreement with data). 

The model we chose is simply too crude and thus the inconsistency is extreme, reflecting the fact that the model is missing some of the important physics of the problem, and the simple physical mechanisms that are retained in the model are incapable of describing the observations with reasonable values for the model parameters. It is not hard to see that such inconsistency of sensitivity is also 
shared by the more complex vortex force model. However, stochasticity can be introduced to tame the inconsistency, as will be shown in what follows.

\section{A Stochastic Balance Model for Longshore Currents}
\label{balance}

The goal is to produce a simple, consistent in sensitivity, balance model for longshore currents. This will be accomplished by enhancing a balance model via stochastic parametrization in order to account for the short-time variability observed in the data. The short-time variability cannot be captured by
the original balance model (\ref{eq:LongHig2}), even when the forcing is obtained from the 8m water observations. Clearly, a great deal of physics is missing. We will only pursue stochastic parametrization rather than to complement the crude model with better or more physics. 

We will begin with an analysis of the data. Two data sets will be used in the model formulation. The ``8m offshore" wave data  largely informs $\beta_B$ in (\ref{eq:LongHig2}), and a longshore time series data will be used as ``training data." We will then compare the 
training data, which is the longshore velocity time series at a specific location, to the longshore model driven by the 8m wave data. The parameter $c_D$ in the drag force will be calibrated to enhance the role played by the drag term in (\ref{eq:LongHig2}) in  constributing to the empirical histogram that results from a comparison of the training data and the longshore model driven by the 8m offshore data. The remaining discrepancies in the empirical distributions are then brought to a minimum by the addition of a stochastic parametrization of momentum contributions. The  discrepancy in the histograms of the forcing and the longshore velocity  is presumed largely due to missing  physics in the original balance model. 
A model that  assigns to stochasticity as little of the model fidelity as possible
is preferred, and further,   we  favor simple noise processes rather than   complicated ones.  We will use  make frequent use   of Gaussian mixtures
to capture non-Gaussianity.  See \cite{GM} for details on Gaussian mixtures. 

 \subsection{Analysis of the Data}
\label{data}
We will be focusing on longshore velocity field data from one particular location, station v13. This is the ``training data." We will denote this data time series as $V$, in what follows.
Some of the basic statistics on the $V$ data appear in Table 1. 
\begin{table*} 
 \caption{{\it First row: empirical statistics of the time series $V$, the longshore velocity $v$ (m/s) at station v13. Subsequent rows: the wave quantities, measured at 8m depth offshore location: the period $T$ in seconds, the significant wave height $H_{mo}$, and wave direction $\theta$ (degrees). Duck data,  September, 1994.}}
\begin{center}
\begin{tabular}{|c|c|c|c|c|c|} \hline
 variable & mean & median  & mode & standard deviation\\ 
\hline
$V$  (m/s) &0.0373& -0.0130 &  -0.032 &  0.2931 \\
$T=2 \pi/\sigma$ (s) & 9.907 & 9.706 & 13.56 & 3.114\\
$H_{mo}= 2 \sqrt{2} A_\infty$ (m) & 0.8272& 0.6410 & 0.3850  & 0.5821\\
$\theta$ (degrees)  & -2.160 & -4.000  & -8.000 &  21.15\\
\hline
\end{tabular}
\end{center}
\label{tab:newtab}
\end{table*}
These statistics pose a challenge  to the balance model (\ref{eq:LongHig2}) and it is apparent in Figure \ref{fig:estd}: if we use mean values for
the data $A_\infty$ as well as $k(T)$ and $\theta$ and we insist that the drag force parameter $c_D$ is positive, the model will generate a negative mean estimate for $V$.  One might think that using the more complex model presented in the Appendix will circumvent the conundrum, however, the derivation of the longshore balance mode using the vortex force model as a starting point  makes it plain that the problem would be present had the vortex force model be used.  (One could use the median or modal values for the
quantities in question, but the model appearing in the Appendix is a mean-field model (see \cite{MR99, MRL04})). 

\begin{figure}
\centering
(a)\includegraphics[width = 0.5\textwidth]{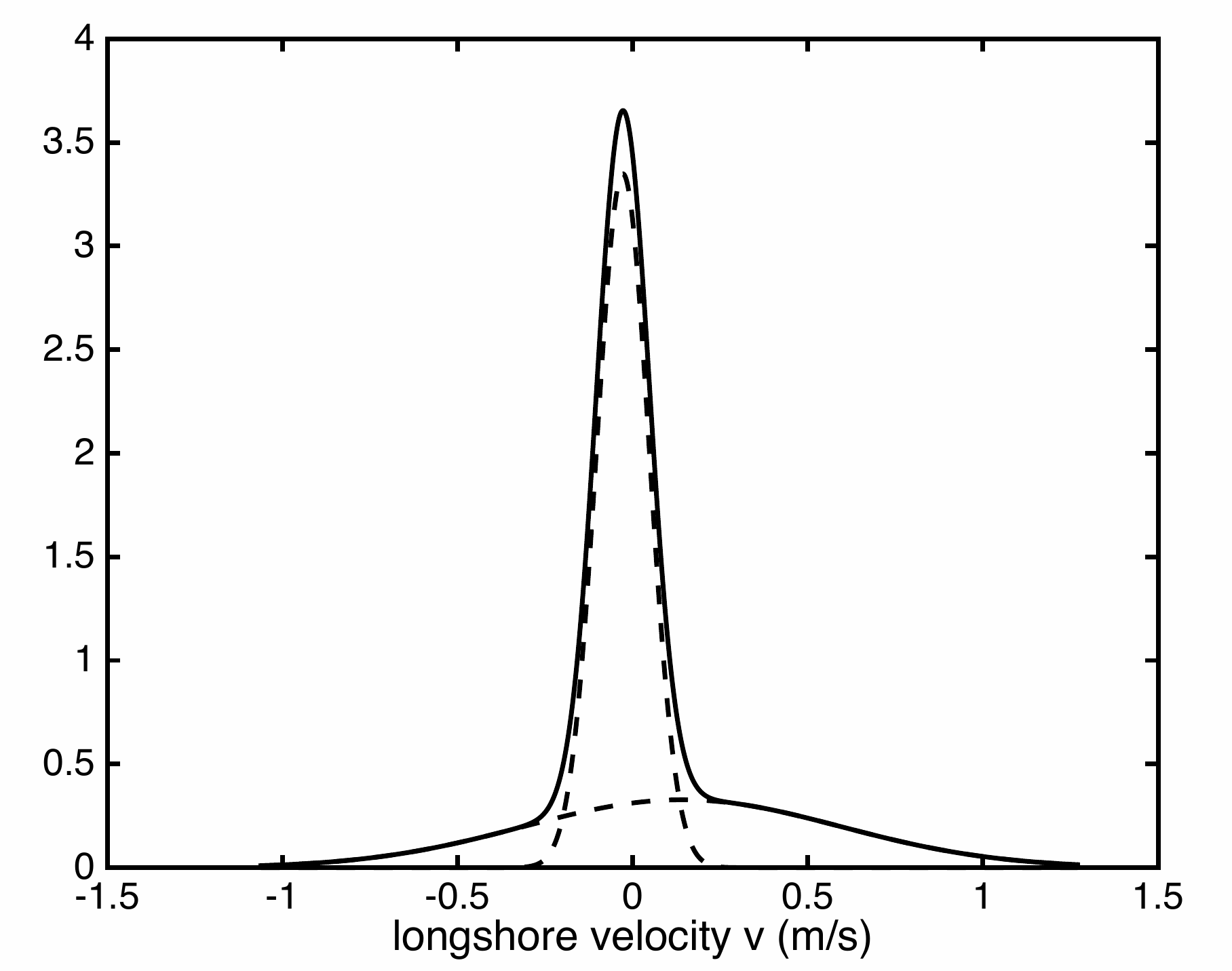}
(b)\includegraphics[width = 0.5\textwidth]{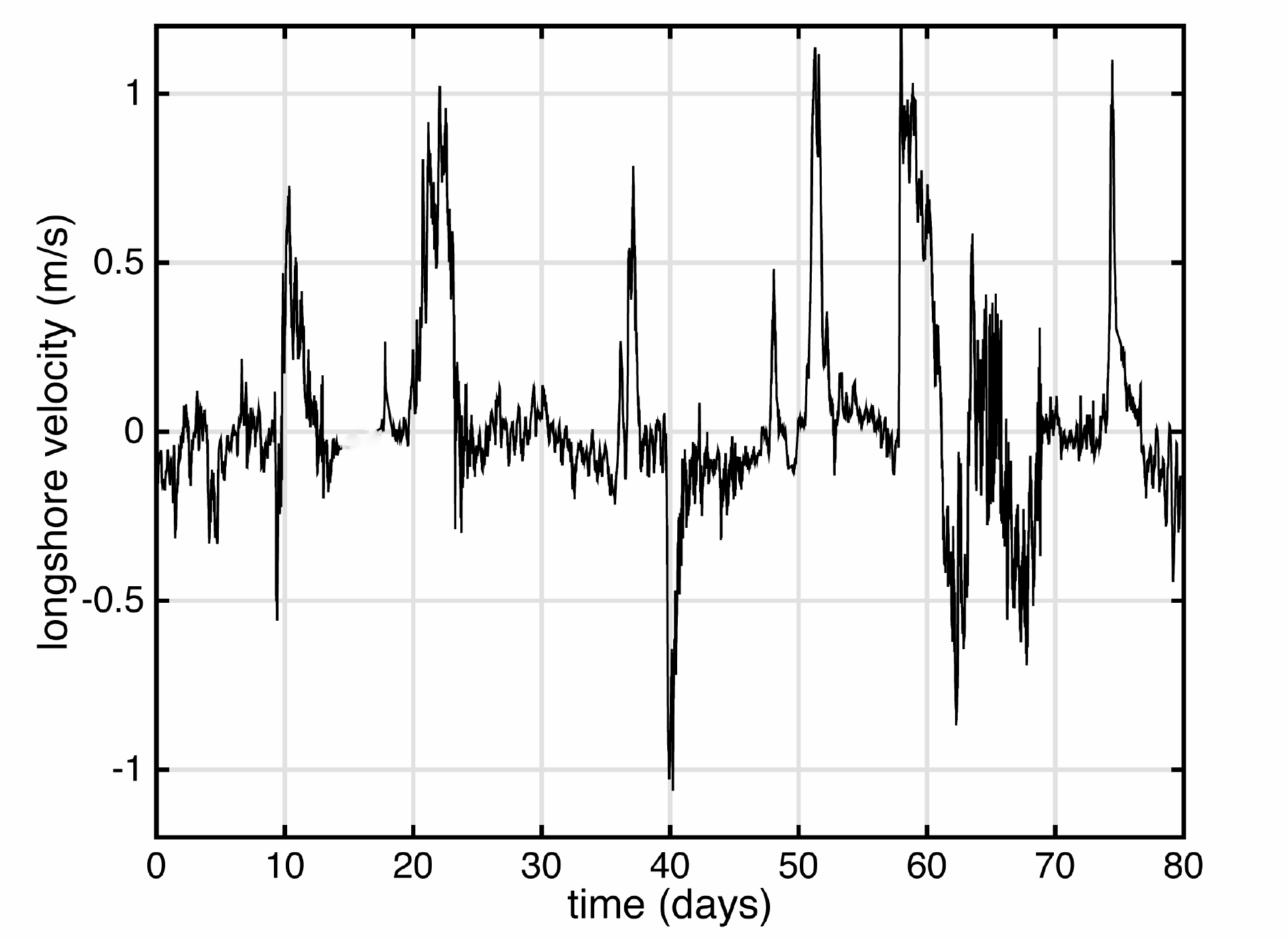}
\caption{Duck data. Longshore velocity $V$.  Time series data at station v13 (measured about 220m offshore), starting on August 1, 1994. (No data is available during days 14-17).
The empirical, Gaussian mixture appears in (a) as dashed curves. Superimposed is the empirical  pdf of the longshore velocity $V$.  
The  distribution is  non-Gaussian, {\it i.e.}, skewed. The
 time series of $V$ appears in (b).
}
\label{fig:v0}
\end{figure}

 We  use instead  an ensemble  modeling approach.
 The  wave forcing will be computed via (\ref{aeq}) using  the 8m depth offshore  wave forcing data. Before constructing the stochastic longshore model we will examine aspects of the statistics of the data that provide valuable physical constraints for the modeling process. 
\begin{figure}
\centering
(a) \includegraphics[scale=0.3]{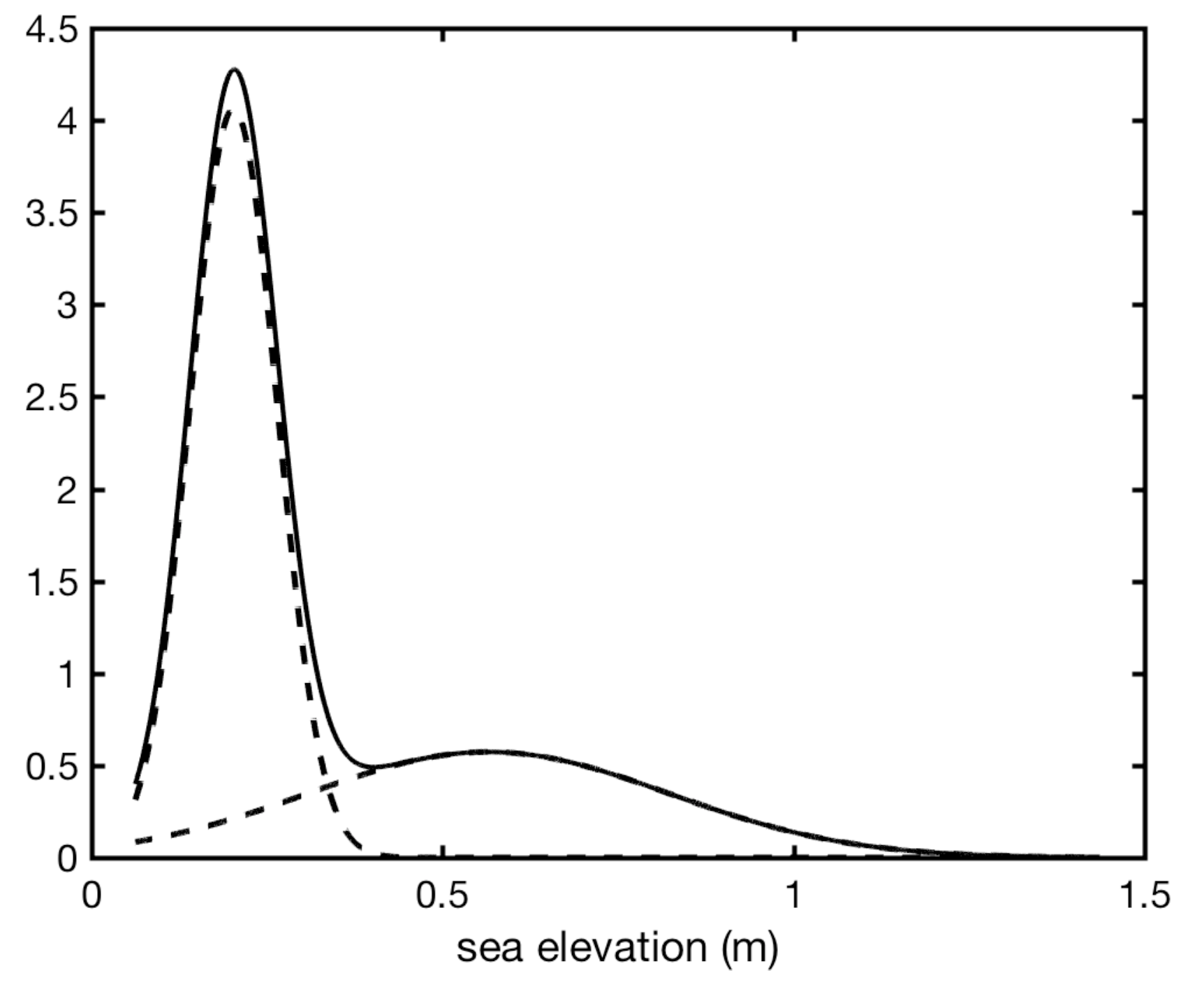}
(b)\includegraphics[scale=0.3]{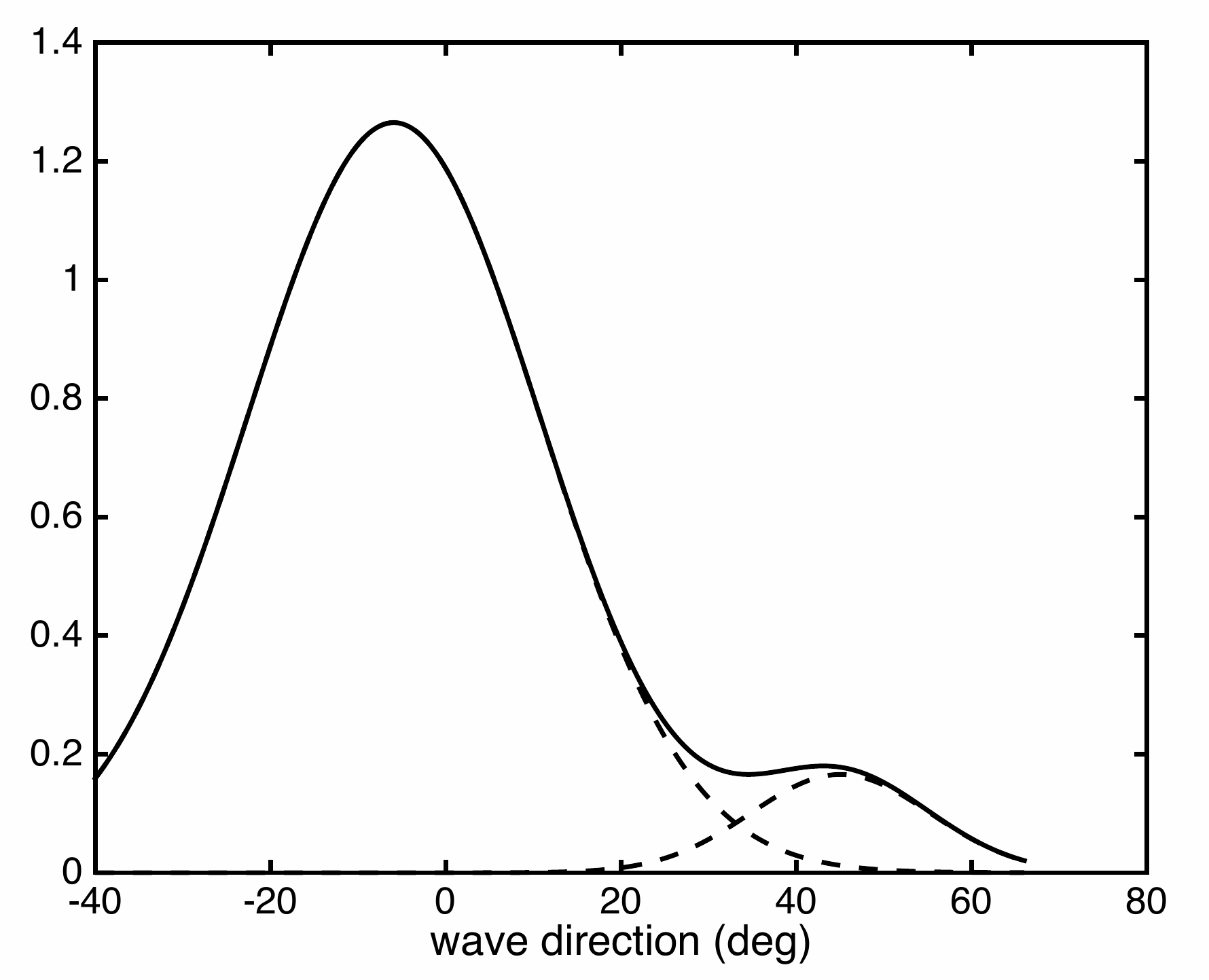}
(c)\includegraphics[scale=0.3]{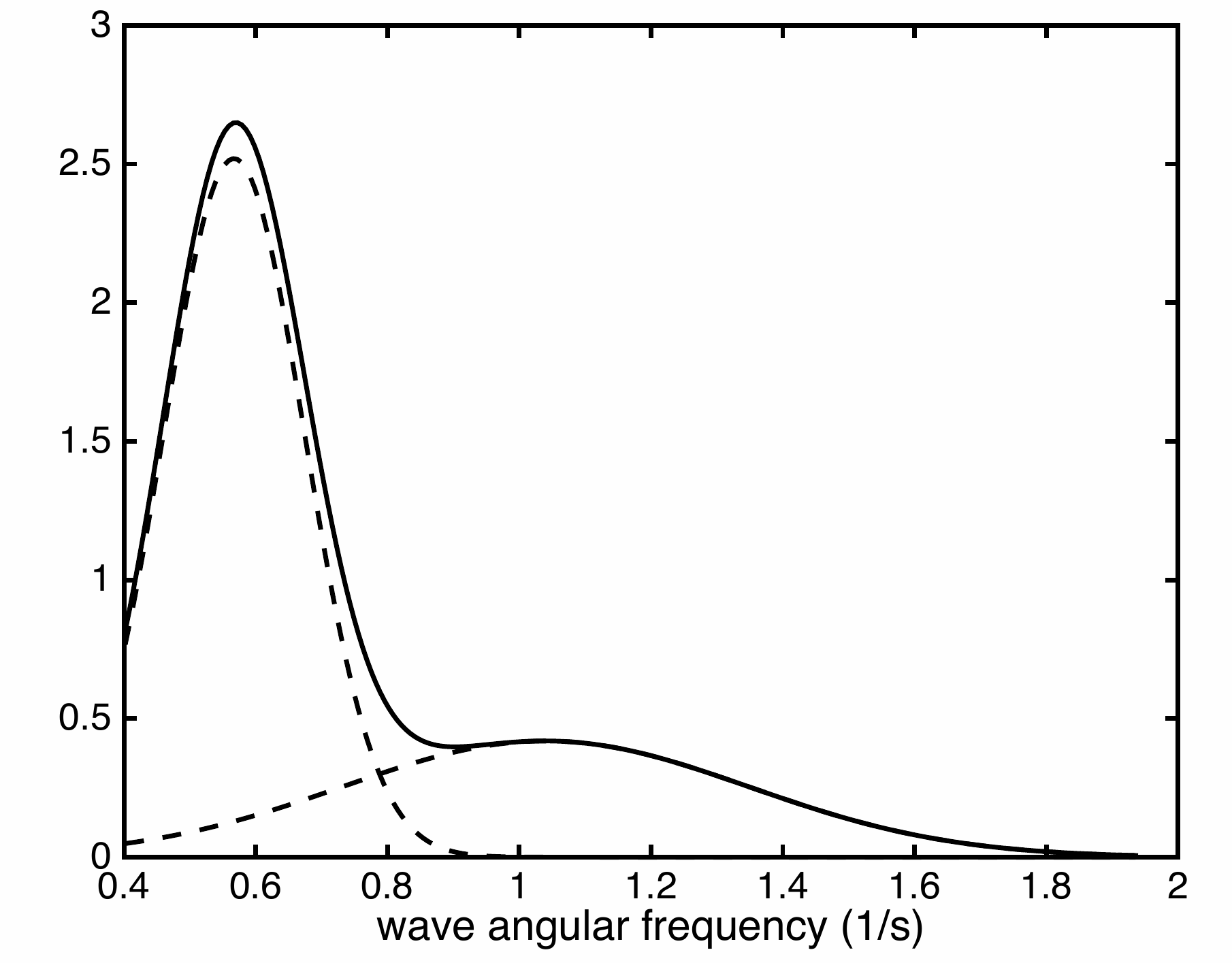}
\caption{Duck wave data (8m depth offshore data). Histograms of the (a) sea elevation (wave height) (m), (b) wave direction (degrees) and (c)  angular frequency (1/s).  The times series was collected at a 2 Hz rate, continuously for 8192 seconds, every 3 hours, during August-October, 1994. Gaussian mixture, solid,
the two Gaussians of the mixture, dashed.
}
\label{fig:hist8}
\end{figure} 
We begin, however, with finding approximations to the empirical probability distribution functions (pdfs) of the data. 
Figure \ref{fig:hist8} shows that  a Gaussian mixture   \cite{CB} approximates   well the pdfs
   of the sea elevation, wave direction, and wave frequency of the 8m depth offshore data. 
 We  specifically use the Expectation-Minimization Algorithm (EM, hereon) in the Gaussian mixture calculations.
   Superimposed, and appearing as dashed lines, are the 2 Gaussians used in the pdf mixture representation. We highlight the skewness of the distributions. Skewness is also evident in the Gaussian mixture approximation of the empirical 
 histogram of $V$. See  Figure \ref{fig:v0}a.  In Figure \ref{fig:v0}b we display the  time series of the longshore velocity $V$, for several months, starting on August 1 and running till the end of October, 1994. This data was used to produce the empirical histogram for $V$. Prominent in the time series is the faster variability occurring at the hourly time scale, as well as the  slower variability changing on  multi-day time scales. 

The connection between the offshore data and the longshore velocity is through the 
term
\begin{equation}
\beta_B(A_\infty,k,\theta,x) = A^7(x;A_\infty) k(x; \sigma) \sin(\theta).
\label{amp}
\end{equation}
 The $\beta_B$ appears in the second term on the right hand side of (\ref{longshore}) and in the second term on the right hand side of (\ref{eq:LongHig2}) (see also (\ref{epsilon})). The term $\alpha \beta_B/h^6$ is a parametrization of momentum exchanges between wave breaking processes and currents.  
The relationship between $k$, the wavenumber and the 8m offshore frequency data  is found via the dispersion relation (\ref{dispersion}). The relationship between the amplitude $A(x)$ and the 8m offshore wave amplitude data  is found approximately as follows:
 assuming  that the variability of the waves is not resolvable at the current time scales, (\ref{eq:waveaction}) is then
\begin{equation}
\frac{d}{dx} [W {\bf C}_g] = -\frac{\epsilon}{\sigma},
\label{reduceda}
\end{equation}
ignoring alongshore variation and taking  $H \sim h$, and the group velocity $|{\bf C}_G| \approx \sqrt{g h}$ (the
group velocity is otherwise found to be given by (\ref{group})). 
The dissipation term $\epsilon$ is defined in (\ref{epsilon}).
The wave amplitude is then given approximately,  by
\begin{equation}
A(h;A_\infty) = h^{-1/4}[ h_\infty^{-5/4} A_\infty^{-5} -\tilde \delta (h^{-23/4}-h_\infty^{-23/4})]^{-1/5},
\label{aeq}
\end{equation}
with $\tilde \delta = \frac{10 \delta}{23 \beta}$. See \cite{TG86}. Here, $\delta = 2 \alpha \sigma g^{-3/2}$.  The amplitude of the waves, or wave forcing,  is $A_\infty$, at depth $h_\infty:=\beta x_\infty$.  The
 depth $h_\infty \approx \lambda/20$, where $\lambda$ is the wave wavelength. This is a depth at which there is less than a 1\% discrepancy between $H$ and $h$ in the shallow water wave limit, $kH$  (we note that (\ref{aeq}) can yield unphysical results, as it can lead
 to $A$ being zero for certain values of $h$). The model for $\epsilon$ (see \ref{epsilon})
 could be modified to account for viscous dissipation, which becomes
 more prominent very near to the shore. (In  \cite{r07} it is shown that stochastic  variability meant to capture episodic wave breaking
 leads to a Stokes drift velocity that has the familiar deterministic component as well as a diffusion-dominated term, however, this 
 type of dissipation would be overwhelmed by nearshore breaking).

\begin{figure}
\centering
\includegraphics[scale=0.3]{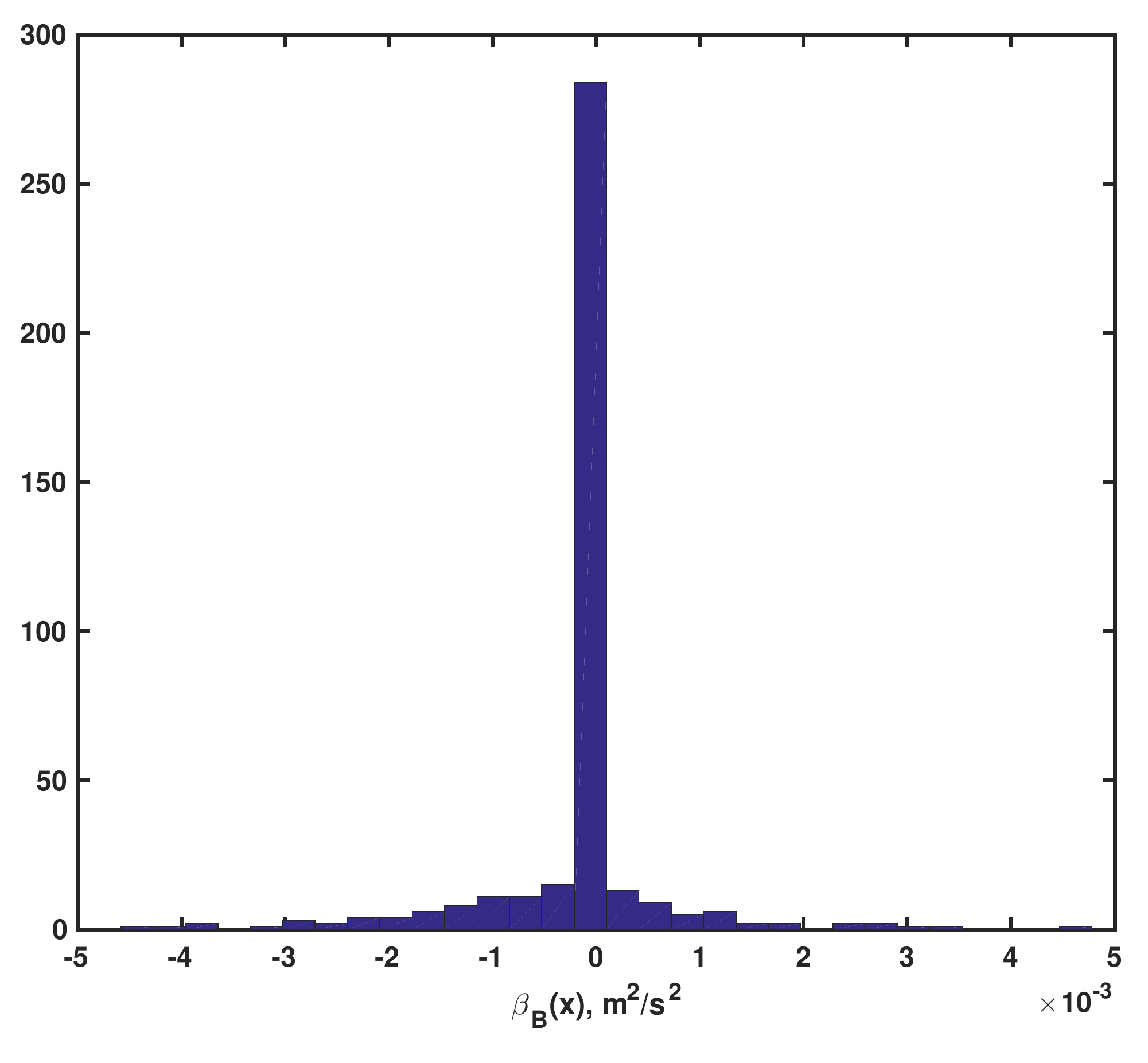}
\caption{Histogram of $\beta_B$. See (\ref{amp}).  Computed by drawing samples of  $k$, $A$, and $\theta$, assuming these
are independent.  Note skewness.}
\label{distbeta}
\end{figure}

In the stochastic parametrization approach we work with ensembles. We first need to create a distribution for $\beta_B$.
Samples of $\beta_B$ can be produced via rejection Monte Carlo sampling (see \cite{rMC}), using the 8m offshore data. 
 The empirical historgram that results from the rejection sampling using the offshore data appears 
in Figure \ref{distbeta}. This  distribution was obtained under the assumption that the samples from $A$, $k$ and $\theta$ 
were independent. We note that the distribution obtained this way is skewed and very narrow (it is even narrower and similarly skewed when the assumption of independence is not used).

\begin{figure}
\centering
(a)\includegraphics[scale=0.3]{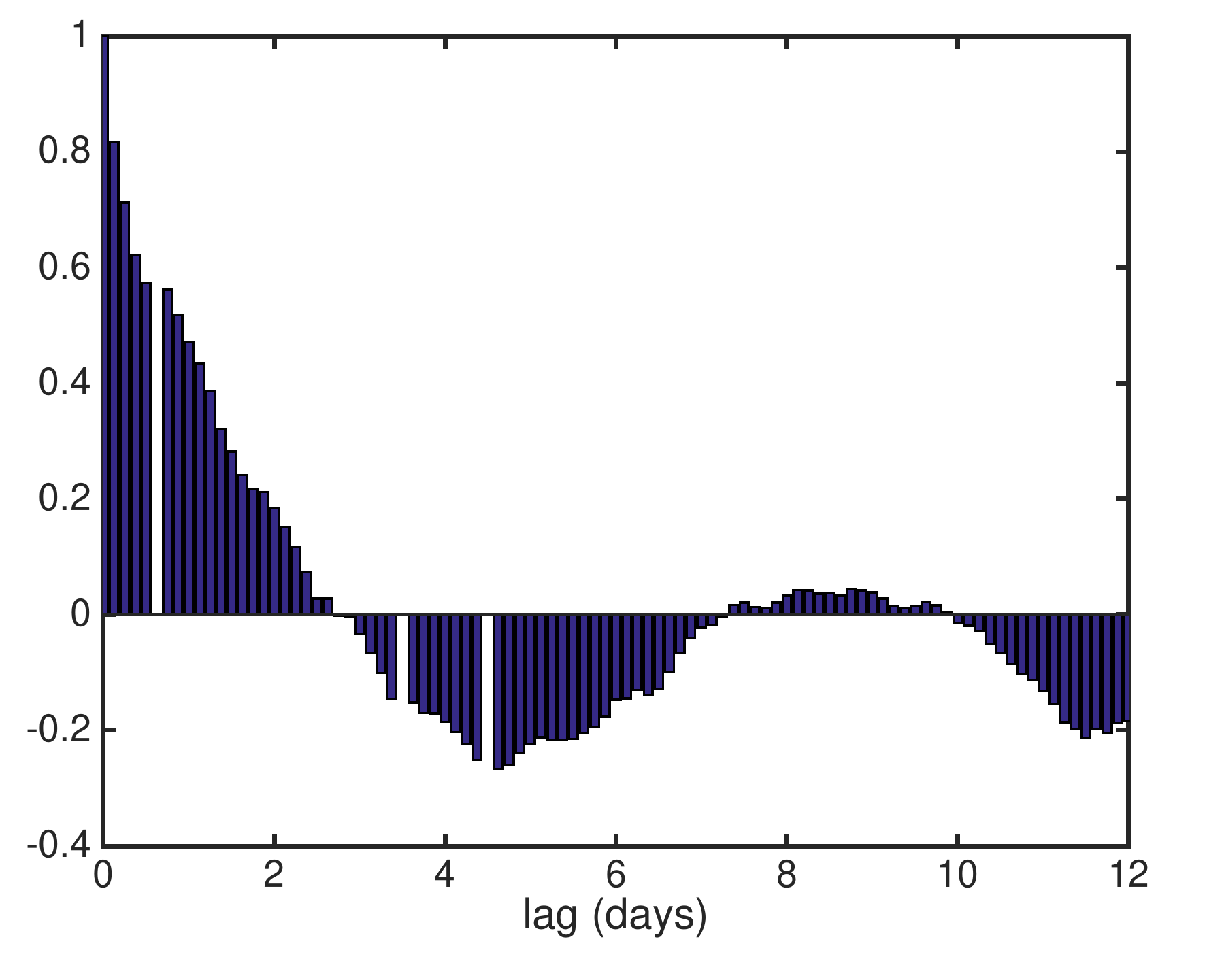}
(b)\includegraphics[scale=0.3]{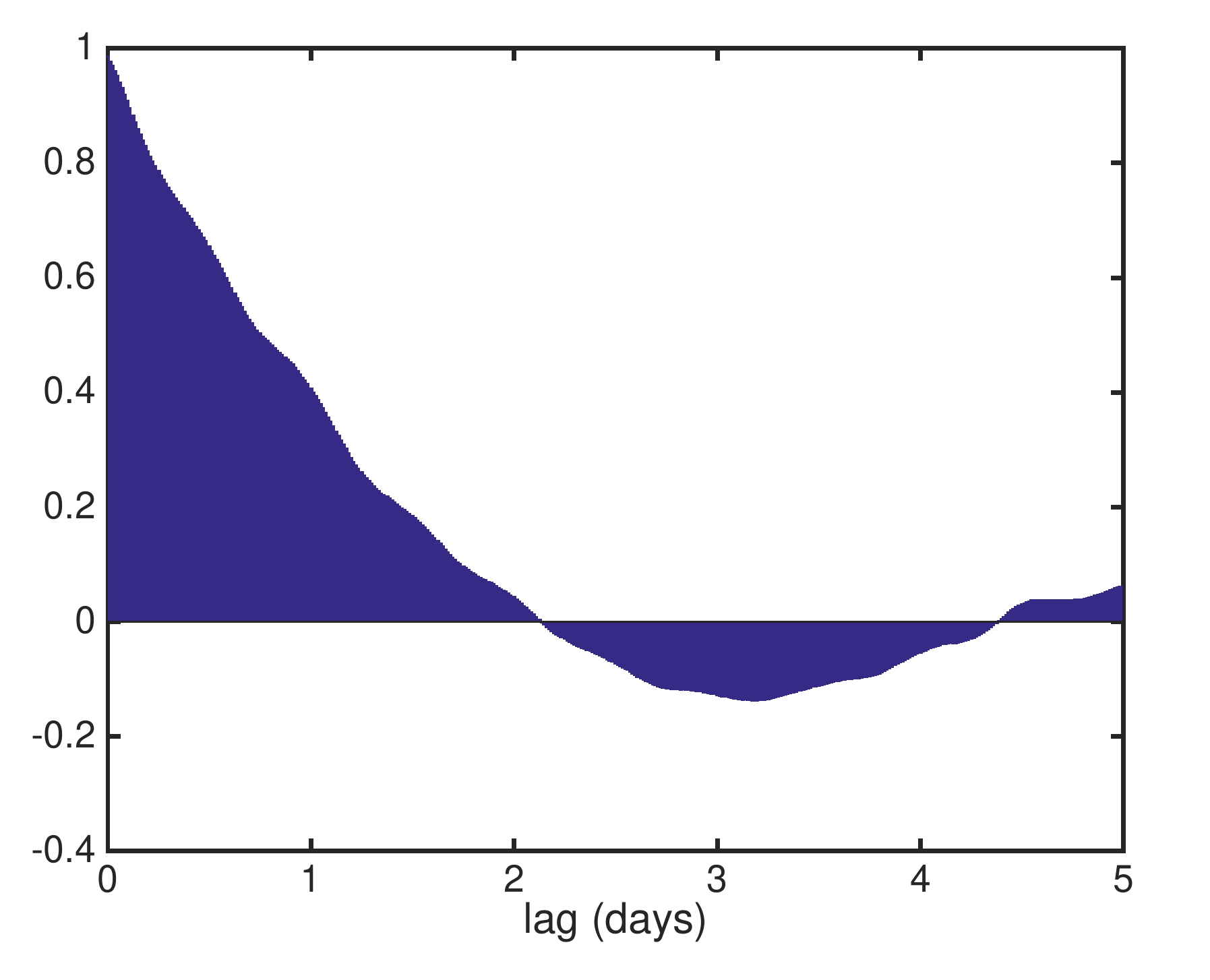}
\caption{(a)
The autocorrelation $\tau_\beta$ of $\beta_B$, using  $A(A_\infty),k,\theta$  observations as input. We denote this the  autocorrelation time $\tau_{\beta}$ and it  is about 1 day.  It is found by calculating a time series of $\beta_B$ from time-ordered triplets of $A(A_\infty),k,\theta$  observations. In (b) we display the autocorrelation $\tau_{V}$. This is the autocorrelation of the time series of  the longshore velocity data at station v13. We note that $\tau_\beta$ is
comparable to $\tau_{V}$.}
\label{distbeta2}
\end{figure}

As we will discuss presently,  a comparison of the empirical autocorrelation times of the 8m offshore data and the training set $V$ will play a critical role in assessing  the resulting stochastic longshore balance model.  Figure \ref{distbeta2}a portrays the autocorrelation $\tau_\beta$ of time-ordered $\beta_B$, which found using the 8m depth offshore wave data as input.  The correlation time $\tau_{\beta}$ is about a day. 
It is comparable to the autocorrelation time $\tau_V$, which corresponds to   the longshore current time series data at location v13 (see Figure \ref{distbeta2}b). There are 3 important points to make in connection with the results portrayed in Figure \ref{distbeta2}: the autocorrelation length is similar in the stochastic model and in the data; the autocorrelation length is mainly set by the wave forcing; the stochastic model rightly delivers the autocorrelation; it was not tuned or manipulated to agree with data.

 \subsection{Formulating an Uncertainty Model  via Stochastic Parametrization}
\label{parametrization}

There are 4 terms in the momentum balance, (\ref{longshore}).  As we noted in the previous section, 
the variability of  empirical distributions of $V$ and of the term $\alpha \beta_B/ h^6$ are very disparate. The bottom drag coefficient $c_D$ is tuned to obtain a certain balance with the term $\alpha \beta_B/h^6$.  We will use tune $c_D$ so that the support of  the empirical distribution of $c_D V$ and $\alpha  \beta_B /h^5$ are comparable. The estimate for  $c_D \approx 0.007$ m/s. 

The time series for $V$ is suggestive of at least two time scales in the dynamics. 
The drag term $c_D v$  and the momentum exchange $\alpha \beta_B/h^5$ term, are meant to dominate the momentum balance at 
the longer time scales.  The computed  time autocorrelations shown in Figure \ref{distbeta2} in fact support the claim that there are two very disparate time scales,
 $h/c_D \ll \tau_{\beta}$. 
These disparate time scales provide reasonable justification to lump all of the short-time missing physics, collectively represented by 
$\partial v/\partial t$, by an additive stochastic term. 
In \cite{noyes1, noyes2}  this short term variability is ascribed to shearing instabilities (see \cite{jsallen}). Shorter time scale  wave breaking
is also noted as a significant fast-time influence on longshore currents (see  \cite{cienfuegos}).    
  We thus proceed with the stochastic parametrization, focusing on capturing the missing fast-time variability. 
  
\begin{figure}
\centering
\includegraphics[scale=0.5]{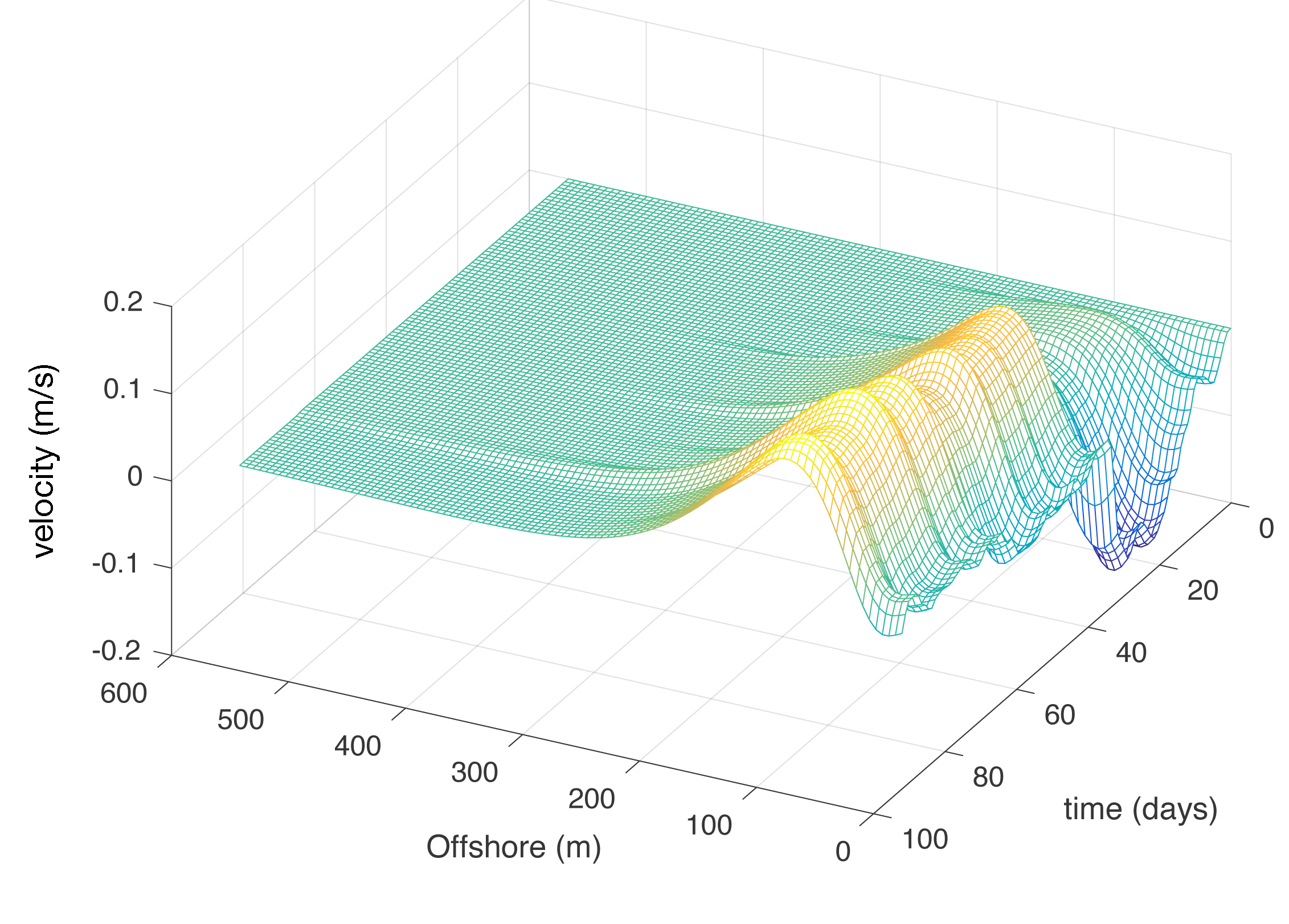}
\caption{Space-time plot of $v_0(x)$.}
\label{full}
\end{figure}

  First, 
since the mean of the data-informed $\alpha \beta_B/ h^6$ and of $c_D V/h$ are not
 the same, the balance of these two terms will on average be non-zero, leading to  a steady acceleration and thus a bias in the longshore current.
In view of these constraints  we  propose a stochastic model for the predicted velocity $v(x,t) := \langle V \rangle + v_0(x) +  v'(x,t)$, where 
$\langle V \rangle$ is the data mean. 
  The space-time solution of $v_0(x)$ appears in Figure \ref{full}. 
  The {\it stochastic longshore velocity balance model} we propose (which for completeness includes the lateral dissipation)  reads
\begin{eqnarray}
0 &=&  ({\cal L}- \frac{c_D}{h})   v_0 - \alpha \frac{  \langle \beta_B \rangle }{h^6} + \alpha \frac{ \beta_B  }{h^6} -  \frac{c_D}{h} v'(x,t,\eta_t),  \label{sde1} \\
v'(x,t,\eta_t) &= &   \sum_{\ell=1}^{N_\ell} f_\ell(x,t) P_\ell(\eta_t) . 
\label{sde2}
\end{eqnarray}
 ${\cal L} w= N  \frac{1}{h}  \frac{\partial}{\partial x}\left(\sqrt{g h} h  \frac{\partial w}{\partial x} \right)$.
 $\eta_t$  is a random variable with known associated
 distribution $\rho(\eta_t)$.
$P_\ell$ are (in this instance)  Hermite  polynomials of degree $0 \le \ell \le N_\ell$. The expansion coefficients $f_\ell$ are found via standard projection, since $\int_{-\infty}^\infty P_\ell(\eta_t) P_j(\eta_t) \rho(\eta_t) d \eta_t = \sqrt{\pi}\delta_{\ell,j}$. The Gaussian measure $\rho(\eta_t)$ is obtained from observational data. The Gaussianity of this measure is assumed and it represents measurement
"error."
See  \cite{polychaos1} for further details.

\begin{figure}
\centering
\includegraphics[scale=0.5]{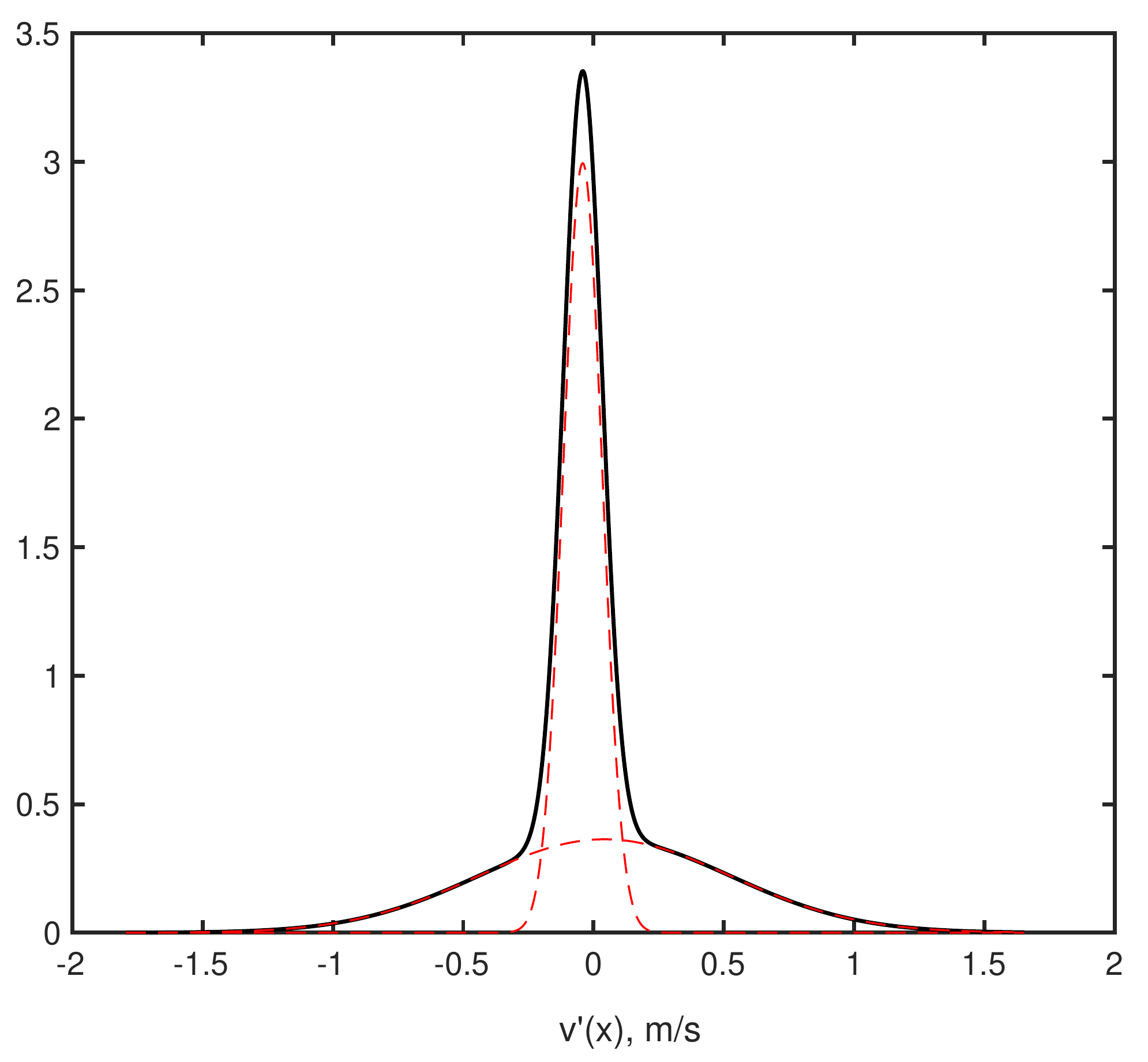}
\caption{Empirical pdf for $v'$  (see also Figure \ref{fig:v0}a).}
\label{outcome}
\end{figure}

Since the model in question is simple we will dispense with the polynomial chaos representation in favor for a
crude Gaussian mixture parametrization of $v'$.
Specifically, 
\begin{equation}
v'(t) =   \sum_{\ell=1}^{N_\ell} a_\ell  {\cal N}_t(m_\ell,\sigma_\ell(v_0)), 
\label{sde3}
\end{equation}
${\mathcal N}_t$ are Gaussian variates. $v'$  can  have a non-zero mean.    
   The input to the EM is  the data-informed quantity $-c_D(V-\langle V \rangle) + \alpha/ (\beta_B - \langle \beta_B \rangle)/h^5$, at  $x$ corresponding to station  v13. For the data under consideration $N_\ell=2$ was adequate.
An empirical histogram of  $v'$ is shown in Figure \ref{outcome}. Using dashed lines we highlight the 2 Gaussians in the mixture. In what follows we will 
assign $\ell=1$ in (\ref{sde3}) to the narrow Gaussian in the mixture, and label the broader one by $\ell=2$.
\begin{figure}
\centering
\includegraphics[scale=0.4]{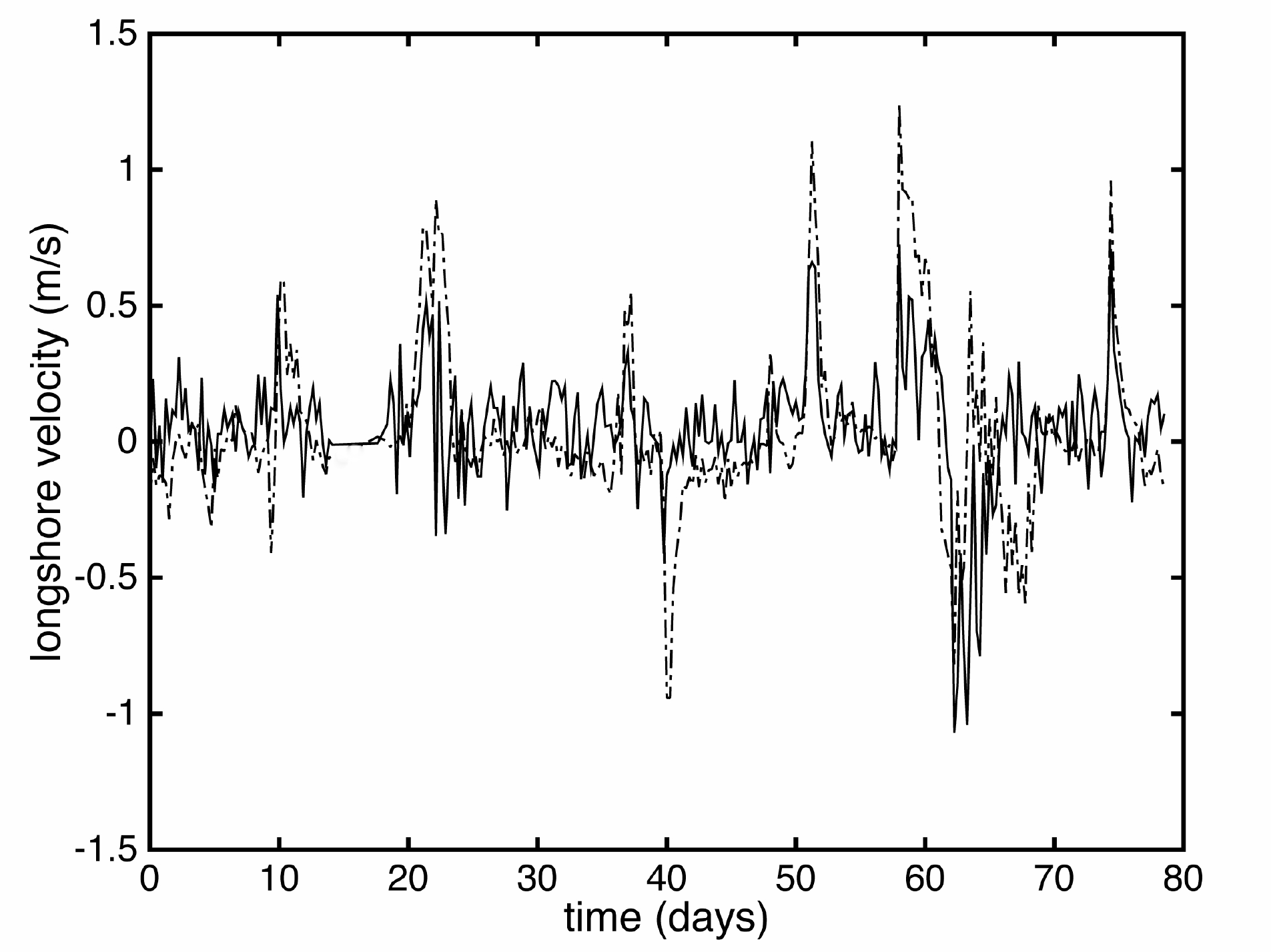}
\caption{Time series of $V-\langle V \rangle$ (dashed) and model outcome $v(x,t)-\langle V \rangle$ at $x$ corresponding to station v13 (solid).}
\label{outcome2}
\end{figure}
The Gaussian mixture captures the empirical pdf  well, and  the model is complete. 

In the next section we will apply the model  in a data assimilation exercise. Stochastic models are well suited to Bayesian inference 
since they can provide concrete priors. In anticipation of this application, we will actually suggest a less accurate but considerably simpler 
representation for $v'$.  It  is clear that the mixture component centered at the mode, the $\ell=1$, has very narrow variance and could be approximated
by its mode. We retain the $\ell=2$ Gaussian mixture component.
 The  stochastic  longshore balance model, simplified,  is thus
\begin{equation}
 v(x,t) \approx - \frac{\alpha}{c_D}  \left[ {\cal L}  -\frac{c_D}{h} \right]^{-1} \frac{ \beta_B(x)}{h^6(x)}
  - a_1 + a_2 {\cal  N}_t(m_2,\sigma_2).
\label{themodel}
\end{equation}
where  $a_1 = 0.04$, $a_2=0.4$,  $m_2 = 0.04$,  $\sigma_2 = 0.4$. 
Adjusting for means, Figure \ref{outcome2} superimposes $v(x)$ at station v13, given by (\ref{themodel}), and $V$. 

\section{Data Assimilation using the Stochastic Balance Model}
\label{da}

We apply the stochastic  longshore balance model in a data assimilation setting of longshore current data. (See \cite{wunschbook}, for
background on data assimilation, with an emphasis on  geophysical applications).  Since the model (\ref{themodel})  is linear and Gaussian, an optimal estimator is found via least-squares, or sequentially, via Kalman filtering. We opt for the Kalman filter. The Kalman  estimator minimizes the trace of the posterior, time dependent conditional probability distribution of  the longshore velocity.  The Kalman filter 
will deliver the time dependent ensemble  mean and variance of the posterior distribution of the longshore velocity, given observations, taking into account model error and observational errors, both assumed known.

To clarify what this simple illustrative calculation demonstrates, we mention the following: the {\it state variable} to be estimated is the time history of the longshore velocity at some location, given observations of the longshore velocity at that location. In order to make the problem more interesting we 
chose to estimate the longshore velocity at station v14, which is about 20m further offshore than location v13. Let $V_{14}$ denote the velocity at station v14. The ingredients in this exercise are, (1) the actual data $V_{14}$, which is to be called {\it truth} and used to assess how good our estimate is; (2) 
the  {\it model} which is a simple discretization of the stochastic longshore model, evaluated at station v14; (3) the 
{\it  data}, which  consists of noisy longshore data $V_{14}$, at  regular time  intervals. (The noise is meant to emulate measurement error. Obviously, the station data comes equipped with measurement errors, however, the inherent variability of the signal does not reflect the measurement error itself). 

The goal is to show that the assimilation of data and the model produces a reasonable estimate of the ensemble mean and variance of the distribution of $V_{14}$. Note that the data $V_{14}$ is highly non-Gaussian in its distribution and thus higher order moments are not going to agree. Moreover, this estimate should be better than that obtained by the model alone or the observations alone. The quality of the estimate is evaluated here by
a simple comparison to the truth data, and expect that a good estimator is close to the truth. 
The model is stochastic, and as suggested in Figure \ref{outcome2}, well trained to capture the data at v13. If we use the model only we will produce ensemble members that resemble those of Figure \ref{outcome2}. At v14, the model yields a mean velocity estimate of
$ \langle v$(v14)$\rangle = 0.0345$ m/s. The mean of the truth is $\langle V_{14} \rangle=0.0311$ m/s. The model does reasonably well.
If we were to use the data only to estimate truth, it will largely depend on how often we measure it and how large the measurement uncertainty is.
Next we consider the data assimilated estimate.

\begin{figure}
\centering
\includegraphics[scale=1,width=3.4in]{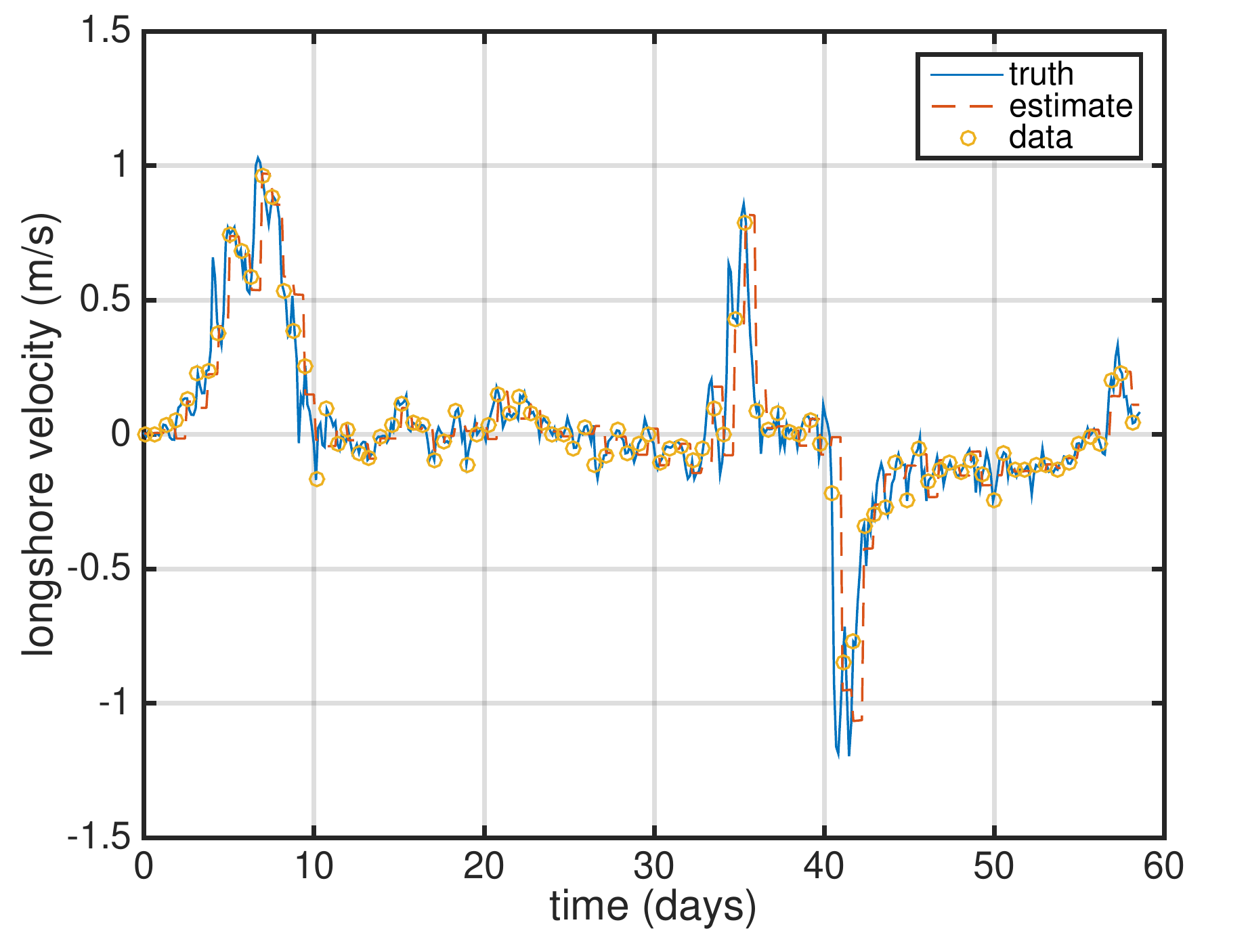}
\caption{Estimation of the longshore velocity data (dashed) using a Kalman Filter and a simple discretization of the longshore model (\ref{themodel}). The
{\it truth} signal is field data (solid).  The measurements  (dots) are read every 5 time steps. The measurements has been synthetically perturbed by a normally-distributed process with variance 10 times greater than the 
variance inherent in the stochastic drag parameter.}
\label{fg.kf}
\end{figure}

Let $\phi_n$ be an approximation of the longshore data at time $t_n$. The discrete model for the longshore current and the relationship between
the model variable $\phi_n$ and the observations $\{y_m\}_{m=1}^M$, respectively,
\begin{eqnarray*}
\phi_{n+1} &=& p \phi_n + \Delta t q_n + B \Delta W_n, \quad n=0,1,2...,  \quad \mbox{the model},\\
\phi_m &=& y_m + C \Delta W_m, \quad m=1,2,..,M, \quad \mbox{the data.}
\end{eqnarray*}
The discrete model is obtained by
using the simplest possible discretization  of  (\ref{themodel}).  Here, $p = 1-\Delta t c_D/h$, and $q_n = \Delta t \alpha k \sin \theta A^7/h^6$. We take $\phi_0 = -a_1+ a_2 m_2$, and $B=a_2 \sigma$.  The dissipation term will be  omitted.  In what follows we will be  assuming that the discrete model  times $t_n$ and  the data times $t_m$  are commensurate. Also, we will assume that  $\Delta t= t_{n+1}-t_n$ is constant for any $n$. Here $\Delta W_n$ and $\Delta W_m$ are uncorrelated normal increments with variances $B$ and $C$, respectively, assumed time independent. $C$  is set to $0.1 B$.
The Kalman filter formulas are
\begin{eqnarray}
\hat \phi  &=& p \hat \phi_n + \Delta t q_n, \quad n=0,1,2,... \nonumber \\
\hat u &=& p \hat u_n p + B_n, \quad n = 0,1,2,... \nonumber\\ 
K_n &=& \hat u(\hat u + C_n)^{-1}, \quad  n = 0,1,2,... \nonumber\\
\hat \phi_{n+1} &=& \hat \phi + K_n(y_m-\hat \phi) \delta_{m,n},  \quad m = 0,1,2,... \nonumber\\
\hat u_{n+1} &=& (1-K_n) \hat u, \quad n=0,1,2,....
\label{kfilt}
\end{eqnarray}
where $\hat \phi_n$ and $\hat u_n$ are  estimates of the mean and the uncertainty, respectively, of $v_n$.
 The ''truth'' is taken as the noise-free field data, which is a 60 day record of the longshore current at station v14, during September and October 1994. Figure \ref{fg.kf} shows the truth, the estimate and the data, $y_m$, which is read every 5 time steps. The variance $u_n$ of the posterior is estimated to be appproximately 0.035, which compares favorably with  the (truth) data variance, estimated at  $0.05$.
We conclude that the assimilated product stays close to the truth and further, that it is statistically consistent with it.

\section{Summary}

Parametrization usually refers to the application of empiricism rather than "first principles" 
in the formulation of some  aspect or the totality of a model of some physical process. By first principles we mean fundamental laws of physics, or conservation or axiomatic principles (such as the least action principle).  Parametrization is used a great deal in problems involving multi-physics and multi-scales and/or as a way to include an important phenomenon in a model that has not yielded to first principle explanations or that cannot be resolved.
The forces in the longshore balance model are in fact   empirically determined,  even though it has a basis in conservation principles of mass, momentum, and energy.   Stochastic parametrization produces a  type of model that makes sense in the  ensemble. The aspect that is parametrized in the model generally does not have a physical or rational basis, however, its inclusion in the model is critical to its fidelity. By fidelity we mean that it is consistent with data in an ensemble sense. Unlike stochastic emulators, which are also aiming for fidelity, and in which structure has to be 
derived from the data and explicitly added, a good stochastically  parametrized model leaves to chance as little as possible and derives its structure from 
whatever physics is being captured by the model.

In this study we purposely picked a  simple deterministic model to use to illustrate how one goes about the task of parametrizing missing physics via 
stochasticity.  We could have instead used a more sophisticated model for longshore currents. 
For 
 example one could start with (\ref{sde1})- (\ref{sde2}), with (\ref{sde1}) replaced by (\ref{eq:momentum}), or perhaps use a wave-resolving model 
such as funwaveC (\cite{feddersen14})). The parametrization process would proceed  in the same way.
The simple longshore balance model, however,  is simply understood and the role played by stochastic parametrization better understood and appreciated. 

 In our specific example the stochasticity was used to represent fast variability that was not resolved by the 
deterministic longshore balance model. A successful outcome of the parametrization is that the stochastic longshore model was able to 
display correlation times similar to those estimated from the data itself.

Situations when stochastic parametrization may be a viable modeling option abound.  As we showed, one such situation is when unresolved but essential physics are important to the faithful representation of observations via models. Other situations introduce stochasticity into model parameters.
One would like to ascribe as much of the phenomena being modeled to physically-based constructs, however, in some instances one is willing to exchange 
rationality for model robustness, or replace the notion of determinism for one in which ensembles make sense.

With regard to model robustness, stochastic parametrization may be a way to achieve better  consistency of sensitivity. Models, particularly ones that have an overwhelming dependence on parametrization often have very delicate tuning of parameters and very limited ranges of physical relevance. Consistency of sensitivity is a property that allows one to assess whether the sensitivity of a model  outcome to changes in parameters is similar to those found in the natural problem
(in the event that the parameter is a physically-meaningful quantity). The consistency of sensitivity analysis   could thus be used as a way to determine the parameter ranges expected for the expected spectrum of physical outcomes.  A model that is consistent in sensitvity will be applicable to a wide variety of physical situations, using sensible or physically meaningful parameter combinations; conversely the analysis can also suggest when this is just not possible. 
Finally,  consistency of sensitivity could be used to compare different models or as another tool for empirical analysis.

 Practically speaking one can see how the introduction of stochasticity can make a model consistent in its sensitivity, at the expense of 
introducing uncertainty/randomness. Our claim is that it is sometimes a reasonable price to pay, especially in complex models of multi-scale/multi-physics phenomena.
Furthermore, the introduction of stochasticity, provided it does not  lead to a serious loss of rationality, can be exploited in a Bayesian setting, wherein forecasts are replaced by ensemble estimates and combine the stochasticity of the model itself and the uncertainties of observations via Bayesian inference and data assimilation.

\appendix

\section{The Vortex Force Wave Current Interaction Model}
\label{model}
Non-wave resolving
  shallow-water  models that capture longshore currents, forced by steady waves and no wind, have 2 known solution manifolds: steady longshore currents and unsteady ones  (see \cite{UMR} and references therein). The stability depends on the strength of the bottom drag, namely, when the drag acceleration is prominent, the longshore currents are steady. In what follows we assume that longshore currents are nearly steady and thus
  we require high drag values in the complex model to simulate these.

\begin{figure}
\centering
 \includegraphics[scale=0.3]{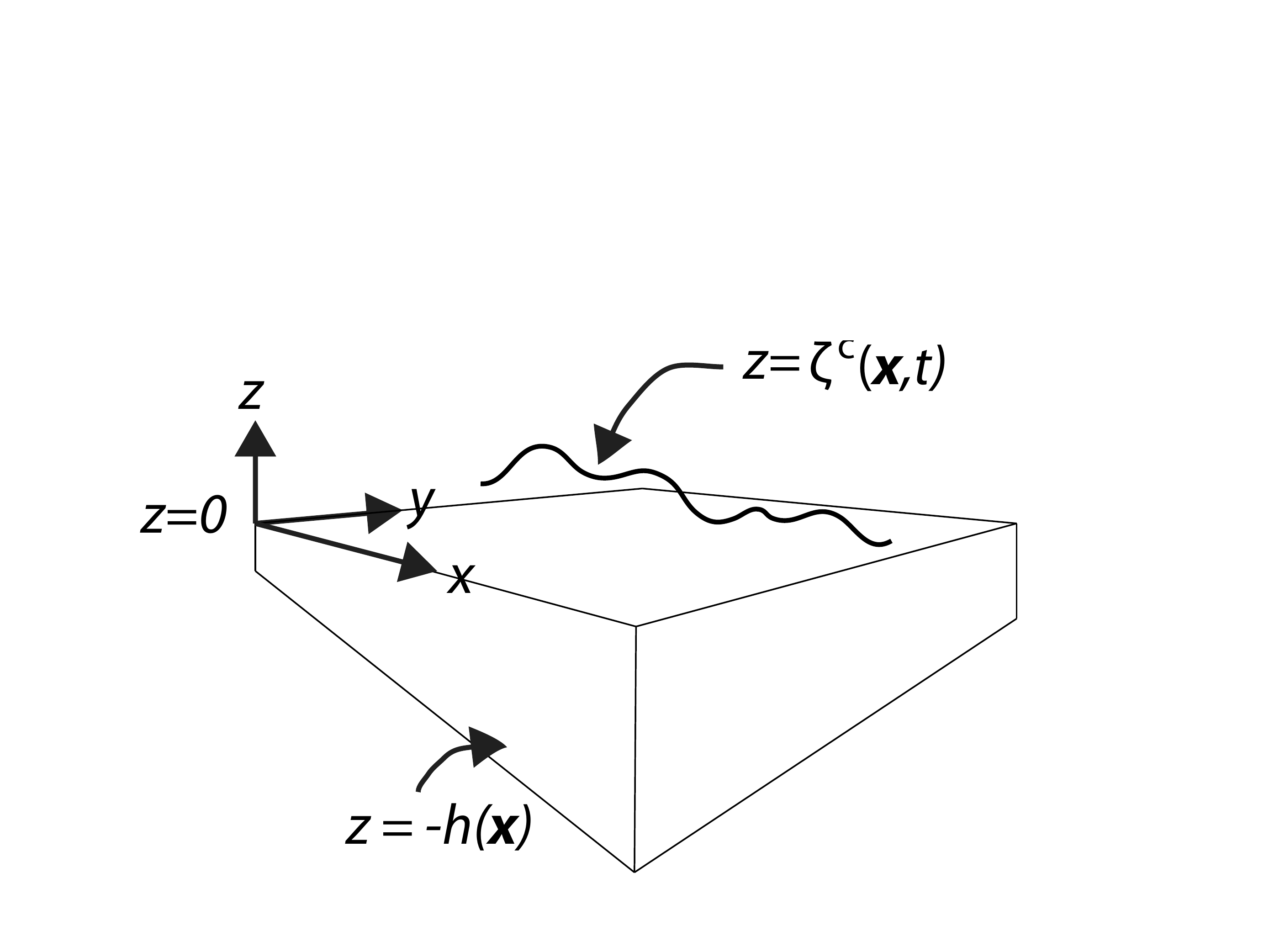}
\caption{Schematic of the nearshore environment, $z$ increases above the quiescent level of the sea $z=0$,  ${\bf x}:=(x,y)$, and $t$ is the (long) time variable. The elevation component of
  the free surface associated with the currents is $z=\zeta^c({\bf
    x},t)$. The
  bottom topography $z=-h({\bf x})$ is referenced to the quiescent sea
    level height, $z=0$.  }
\label{fig:domain}
\end{figure}

The depth-averaged, wave current interaction  model in \cite{MRL04} is specialized to the nearshore 
environment. This is a model for the interaction of currents and waves at  spatio-temporal scales much larger than those typical of the waves.   
A schematic of the domain, along with the coordinate
system is described in  Figure \ref{fig:domain}.
The transverse coordinates of the domain will be denoted by ${\bf x}:=(x,y)$. The cross-shore coordinate is $x$ and increases away from the
beach. Time is denoted by $t \ge 0$. Differential operators depend
only on ${\bf x}$ and $t$. The total water column depth is given by  $H:=h({\bf x}) + \zeta^c({\bf x},t)$,
where
$\zeta^c= \hat \zeta + \zeta$, is the composite sea elevation. The sea elevation has been split into its  dynamic component $\zeta({\bf x},t)$, and $\hat \zeta$, the quasi-steady sea elevation adjustment.
$\hat{\zeta}=-A^{2}k/(2\sinh(2kH))$,
where $A$ is the wave amplitude and $k$ is the magnitude of the
wavenumber ${\bf k}$. The wave frequency $\sigma$  is given by the 
dispersion relationship
\begin{equation}
\sigma^2 = gk \tanh(k H),
\label{dispersion}
\end{equation}
where $g$ is gravity, 
and the evolution of the wave number is found by the conservation equation
\begin{equation}
\frac{\partial\mathbf{k}}{\partial t}+ {\bm \nabla} \omega=0,
\label{eq:wavenumbers}
\end{equation}
where $\omega = {\bf k}\cdot {\bf u} + \sigma$. ${\bf u}({\bf x},t):=(u,v)$ is the depth-averaged velocity (current) vector.
The wave amplitude $A$ is found by solving for the wave action
\begin{equation}
W:=\frac{1}{2\sigma}\rho gA^{2},
\label{action}
\end{equation}
via  the action equation,
\begin{equation}
\frac{\partial W}{\partial t}+\nabla\cdot(W\mathbf{c}_G)  =-\frac{\epsilon}{\sigma},
\label{eq:waveaction}
\end{equation}
where $\rho$ is the fluid density.
The wave action dissipation  rate is captured by
$-\frac{\epsilon}{\sigma}$. The group velocity is 
 $\mathbf{c}_G= \mathbf{u}+ \mathbf{C}_G$, with  $\mathbf{C}_G$ given by 
\begin{equation}
\mathbf{C}_G=\frac{\sigma}{2k^{2}}\left(1+\frac{2kH}{\sinh(2kH)}\right) {\bf k}.
\label{group}
\end{equation}

The  continuity equation is given by
\begin{equation}
\frac{\partial H}{\partial t}+\nabla\cdot [ H(\mathbf{u}+\mathbf{u}^{\mathrm{st}})]  =0,\label{eq:continuity}
\end{equation}
where
\begin{equation}
\mathbf{u}^{\mathrm{st}}:=(u^{\mathrm{st}},v^{\mathrm{st}})=\frac{1}{\rho H}W\mathbf{k},
\label{eq:stokesdrift}
\end{equation}
is the Stokes drift velocity.

The  velocity ${\bf u}$ is found via the momentum equation
\begin{equation}
\frac{\partial\mathbf{u}}{\partial t}+(\mathbf{u}\cdot\nabla)\mathbf{u}+g\nabla\zeta - \mathbf{J}=  
{\bf S} + {\bf N}+ 
\mathbf{B}-\mathbf{D}.
\label{eq:momentum}
\end{equation}
The vortex force term  (see \cite{MR99}) is 
\[
\mathbf{J}=-\mathbf{\hat z}\times\mathbf{u}^{\mathrm{st}}\chi,
\]
where $\chi$ is the vorticity, and $\mathbf{\hat z}$ is the unit vector pointing anti-parallel to gravity. 

The terms on the right hand side of (\ref{eq:momentum}) model several physical processes critical to  nearshore wave-current conditions, none of which have generally agreed-upon  parametrizations: 
${\bf S}$ and ${\bf N}$ represent  the depth-averaged wind stress and sub-scale processes associated with viscous dissipation, respectively.
Wave-to-current momentum exchanges  due to the breaking waves  are captured by
\[
\mathbf{B}=\frac{\epsilon\mathbf{k}}{\rho H\sigma}.
\]
There are several empirical formulations for  $\epsilon$ ($\ge 0$). The one we adopt here
is   due to  \cite{TG83}. (See also \cite{TG86}).  It is
\begin{equation}
\epsilon=24\sqrt{\pi}\rho g\frac{B_{r}^{3}}{\gamma^{4}H^{5}}\frac{\sigma}{2\pi}A^{7},
\label{epsilon}
\end{equation}
with $B_{r}$, $\gamma$, empirical parameters.  This empirical relationship based upon hydraulic theory
and has  been fit and tested against data in nearshore environments similar to the nearshore case
considered in this paper.
The depth-averaged bottom drag is 
\[
\mathbf{D}=\frac{{\bm \tau}}{\rho H},
\]
where 
\begin{equation}
{\bm \tau}= \rho c_D \mathbf{u},
\label{drag}
\end{equation}
where $c_D=c_f |{\bf u}_w|=\frac{2}{\pi} \tilde c_f   |{\bf u}_w|$ is the {\it bottom drag} parameter. $|{\bf u}_w|$ is the wave orbital velocity, estimated near the bottom topography, and $\tilde c_f\ge0$ the friction coefficient.  In this particular bottom drag parametrization the friction coefficient $\tilde c_f$ (or the drag parameter $c_D$ itself) needs to be calibrated/empirically determined.
It is assumed that $|{\bf u}_w|$  is much larger that $|{\bf u}|$ (See \cite{fredsoebook}, Chapter 5) and exhibits  inherent variability at time scales of the waves, which are shorter the variability of the currents. This form of the bottom drag represents the most limited and simplest possible 
parametrization (we will purposely choose this parametrization for this exercise and stochastic parametrization, but other models could be used:
The models of Soulsby \cite{stive1995advances, soulsbybook} and Feddersen and co-workers \cite{feddersen2000} are alternatives, the latter in fact has been tuned to conditions present in Duck NC.  (See also  \cite{UMR} for a comparison of these different drag models).  

\subsection{The Longshore Current Balance Model}

Similar assumptions are made in the derivation of this balance model as were made in connection with 
deriving (\ref{lh}). The longshore current $v$ does not have $y$ dependence. The wave-to-current  momentum transfer
due to wave breaking ${\bf B}$ is retained, as is the ${\bf N}$. Wind stresses are ignored, hence ${\bf S}$ is omitted. 
The crux of the derivation of a balance model for longshore currents, based upon the vortex force formulation of wave-currents, is
the observation that  the  depth-averaged normal component of the velocity at the shore must be zero,
and hence   $u=-u^{St}$. Also for steady currents, the vortex force and the inertial terms balance, provided that the only contribution to the currents are wave-induced, {\it i.e.}, there are no remotely imposed currents or wind stresses to account for. The simple longshore momentum balance will be
\begin{equation}
\frac{\partial v}{\partial t}\approx  - c_D \frac{v}{h} + \alpha \frac{\beta_B}{h^6}  +N  \frac{1}{h}  \frac{\partial}{\partial x}\left(\sqrt{g h} h  \frac{\partial v}{\partial x} \right),
\label{longshore}
\end{equation}
where  
\[
\alpha := 12/\sqrt{\pi} g B_r^3/\gamma^4 \ge 0, \quad \quad
\beta_B:= A^7 k \sin(\theta).
\]
  The wavenumber magnitude is $k$,  
the sea elevation is $A$, and $g$ is gravity. 
$B_r$ and $\gamma$ are parameters associated with wave breaking and sea elevation, respectively.  When we further assume that
$\partial v/\partial t = 0$ in (\ref{longshore}) we obtain the vortex force longshore balance model, (\ref{eq:LongHig2}).

\section*{Acknowledgments}
We received funding from GoMRI/BP. JR and SV also  received funding from 
NSF-DMS-1109856. We wish to thank Prof. Falk Feddersen, for discussions on 
current longshore models.  Several suggestions by the referees improved the 
paper.  JR also thanks the J. T. Oden Fellowship program at U. Texas, Austin, 
and the Aspen Center for Physics. The Aspen Center of Physics is supported, in part, by 
the National Science Foundation under Grant No. PHYS-1066293.

\end{document}